\documentclass[11pt,english]{article}
\usepackage{jheppub}

\usepackage[latin9]{inputenc}
\usepackage{amsmath}
\usepackage{amssymb}
\usepackage{enumitem}
\usepackage{bigfoot}

\usepackage{slashed}
\allowdisplaybreaks
\usepackage[numbers,sort&compress]{natbib}
\usepackage{hyperref}

\hypersetup{
colorlinks=true,
linkcolor=blue,
linktocpage=true,
citecolor=violet
}

\def\Mathematica{{{\sc Mathematica}}}

\def\FiniteFlow{{{\sc FiniteFlow}}}
\def\Lotty{{{\sc Lotty}}}

\usepackage{tikz}
\usetikzlibrary{"arrows", "automata", "backgrounds", "calendar", "chains", "matrix", "mindmap", "patterns", "petri", "shadows", "shapes.geometric", "shapes.misc", "spy", "trees"}
\usetikzlibrary{arrows,shapes}
\usetikzlibrary{trees}
\usetikzlibrary{matrix,arrows} 	
\usetikzlibrary{positioning}				
\usetikzlibrary{calc,through}				
\usetikzlibrary{decorations.pathreplacing}
\usepackage{pgffor}

\usetikzlibrary{decorations.pathmorphing}	
\usetikzlibrary{decorations.markings}
\tikzset{
    fermion/.style={draw=black, postaction={decorate},
        decoration={markings,mark=at position .52 with {\arrow[draw=black]{>}}}},
    fermionbar/.style={draw=black, postaction={decorate},
        decoration={markings,mark=at position .5 with {\arrow[draw=black]{<}}}},
    fermionnoarrow/.style={draw=black},
    phi/.style={dashed, draw=red}
}

\newcommand{\semiloop}[4][]{%
        \draw[#1] let \p1 = ($(#3)-(#2)$) in (#3) arc (#4:({#4+180}):({0.5*veclen(\x1,\y1)});)
}

\makeatother
\usepackage{babel}

\newcommand{\munich}{Max-Planck-Institut f\"ur Physik, Werner-Heisenberg-Institut, 
80805 M\"unchen, Germany.}

\begin{document}
\preprint{MPP-2021-14}

\title{Loop-tree duality from vertices and edges}

\author[a]{William J. Torres~Bobadilla}
\affiliation[a]{\munich}
\emailAdd{torres@mpp.mpg.de}

\abstract{
The causal representation of multi-loop scattering amplitudes, obtained from the 
application of the loop-tree duality formalism, comprehensively elucidates, 
at integrand level, the behaviour of only physical singularities. 
This representation is found to manifest compact expressions
for multi-loop topologies that have the same number of \textit{vertices}. 
Interestingly, integrands considered in former studies, 
with up-to  six vertices and $L$ internal lines, display the same structure of 
up-to four-loop ones. 
The former is an insight that there should be a correspondence between 
vertices and the collection of internal lines, \textit{edges}, 
that characterise a multi-loop topology.
By virtue of this relation, in this paper, we embrace an approach to properly
classify multi-loop topologies according to vertices and edges. 
Differently from former studies, we consider the most general topologies,
by connecting vertices and edges in all possible ways. 
Likewise, we provide a procedure to generate causal representation
of multi-loop topologies by considering the structure of causal propagators.
Explicit causal representations of loop topologies with up-to nine vertices are provided.
}

\maketitle

\section{Introduction}

In the recent years, the calculation of physical observables, relevant for
high-energy physics colliders, turns everyday to be more challenging due to the 
high precision the current experiments are demanding~\cite{Blondel:2019vdq,Banerjee:2020tdt}. 
The Next-to-Leading Order (NLO) revolution, originated by extremely clever ideas,
opened a street, based on formal aspects of scattering amplitudes, 
to provide several phenomenological calculations. 
In effect, the various ingredients that need to be taken into account 
were understood and automated in several frameworks~\cite{Berger:2008sj,Cascioli:2011va,Badger:2012pg,Cullen:2014yla,Actis:2016mpe,Alwall:2014hca,Heinrich:2020ybq}.
Since many of these methods promote 
the space-time dimension from four 
to $d$, new reformulations that aim at 
a local cancellation without altering the dimension are 
yet under study~\cite{Soper:1999rd,Soper:1999xk,Binoth:2000ps,Soper:2001hu,Kramer:2002cd,Becker:2010ng,Becker:2012aqa,Pittau:2012zd,Donati:2013iya,Fazio:2014xea,Primo:2016omk,Mastrolia:2015maa,Hernandez-Pinto:2015ysa,Sborlini:2016gbr,Sborlini:2016hat,Gnendiger:2017pys,Capatti:2020xjc,TorresBobadilla:2020ekr,Prisco:2020kyb}.

In view of the Next-to-Next-to-Leading Order (NNLO) revolution, many ingredients
have to be ready to properly get combined and also allow for a synergy with the
results delivered by the experiments. 
Among the  several obstacles that prevent the NNLO revolution, 
we would like to emphasise the role in the evaluation 
of multi-loop Feynman integrals that is drawing the attention
in the scientific community~\cite{Chawdhry:2019bji,Caola:2020dfu,Kallweit:2020gcp,Badger:2021nhg,Agarwal:2021grm}
where the number of kinematic invariants starts increasing. 
Nonetheless, analytic or numerical evaluation of 
dimensionally regulated multi-loop Feynman integrals with
several internal masses is still a bottleneck that must be sorted out~\cite{Chetyrkin:1981qh,Kotikov:1991pm,Laporta:2001dd,vonManteuffel:2012np,Henn:2013pwa,Argeri:2014qva,Maierhoefer:2017hyi,Borowka:2015mxa,Smirnov:2015mct,Larsen:2015ped,Mastrolia:2018uzb,Abreu:2020xvt,Ita:2015tya}.

In light of providing an alternative approach to the numerical evaluation of
multi-loop Feynman integrals, we elaborate on the loop-tree duality (LTD) 
formalism~\cite{Catani:2008xa,Bierenbaum:2010cy,Bierenbaum:2012th,Buchta:2014dfa,Buchta:2015wna}. 
The LTD formalism, based on the Cauchy residue theorem, 
can be realised as an operation that opens loops into trees and, 
due to the way how it is formulated, it is always possible to integrate out the 
energy component of the loop momenta in any multi-loop
Feynman integral. As a byproduct of this operation, one passes from 
Minkowski to Euclidean space, allowing in this way to compactify all
the physical information in spatial components. 
In other words, a $d$-dimensional Feynman integral in 
Minkowski space, after making use of the LTD formalism, 
is expressed only in terms of the spatial components of 
the loop momenta. 
In fact, the use of the LTD formalism allows to bypass the standard parametrisation
of Feynman integrals in terms of the two Symanzik polynomials
(see~\cite{Bogner:2010kv}  and references therein)
and permits to integrate in Euclidean space.  
Several phenomenological applications by using  LTD, performed
at one and two loops~\cite{Jurado:2017xut,Driencourt-Mangin:2017gop,Driencourt-Mangin:2019aix,Driencourt-Mangin:2019yhu,Plenter:2020lop}, allow for a 
reformulation of this formalism~\cite{Aguilera-Verdugo:2019kbz,Verdugo:2020kzh,Aguilera-Verdugo:2020kzc,Ramirez-Uribe:2020hes,Aguilera-Verdugo:2020nrp}
and motivate alternative studies~\cite{Tomboulis:2017rvd,Runkel:2019zbm,Runkel:2019yrs,Capatti:2020ytd,Capatti:2019edf,Capatti:2019ypt,delaCruz:2020cpc}.

Recently, it has been observed that the application of the LTD formalism 
on multi-loop Feynman integrands leads to a causal representation displaying only
physical information~\cite{Aguilera-Verdugo:2019kbz,Verdugo:2020kzh}. 
In effect, these integrands, written in terms of causal propagators,
allow for a smooth numerical evaluation, avoiding in 
this way numerical instabilities, as discussed in~\cite{Aguilera-Verdugo:2020kzc}.
An alternative approach to our causal representation of Feynman integrands
was also presented in Ref.~\cite{Capatti:2020ytd} by means of H surfaces,
which, within our framework, corresponds to causal propagators. 

Inspired by the compact representation found in Ref.~\cite{Aguilera-Verdugo:2020kzc},
for the multi-loop 
Maximal (MLT), Next-to Maximal (NMLT) and Next-to-Next-to Maximal (N$^2$MLT)
loop topology configurations,
it is desirable to understand the reason of the same structure of the causal 
representation, regardless of the loop order.  
To this end, in this paper, we classify the loop topologies according to the number
of \textit{vertices} and \textit{edges}. For instance, an N$^k$MLT configuration will have $k+2$ vertices,
and the number of edges will correspond to the number of connections between vertices
in a topology. 
Besides, we elaborate on the most general topology generated from a given number 
of vertices and all possible edges. 
The motivation of doing so is to obtain the simplest and most compact 
formulae for the causal representation of the loop topology,
in such a way that topologies made of less number of vertices and edges
can be generated from the latter one by applying certain procedures. 

Furthermore, to relate loop topologies through vertices and edges, 
we introduce two procedures: \textit{removing} and \textit{collapsing}.
The former procedure removes edges by keeping fixed the number of vertices,
whereas the latter collapses a vertex by removing it as well as the edges it is connected.  
These two procedures allows to generate any kind of topology built from
a given number of vertices and allows for a validation of our new results. 
Then, with the pattern displayed by these procedures in the causal representation,
we provide the most general formulae for topologies with up-to nine vertices.
The latter, in the most general case, corresponds to a topology at twenty-eight loops. 

\bigskip This paper is organised as follows. 
In Sec.~\ref{sec:introLTD}, we recap the main features of the LTD formalism,
briefly review the causal representation of the MLT configuration,
and introduce the notation in terms of vertices and edges.
Then, in Sec.~\ref{sec:preliminar}, with the new notation, 
we provide the causal representation of NMLT and N$^2$MLT and
introduce the procedures: \textit{removing} and \textit{collapsing}. 
In Sec.~\ref{sec:fivetopos}, from the knowledge of the former sections,
we generate the most general topology of five vertices and apply the 
removing and collapsing procedures. Besides, we provide a general 
procedure to generate the causal representation of any loop topology.
Then, with the findings of Secs.~\ref{sec:preliminar} and~\ref{sec:fivetopos}, 
in Sec.~\ref{sec:noltd}, we provide causal 
representations of topologies with up-to nine vertices, by only considering 
relations among causal propagators. 
Finally, in Sec.~\ref{sec:conclusiones}, we draw our conclusions
and discuss the future research directions.

\section{Loop-tree duality formalism}
\label{sec:introLTD}
\subsection{Notation}

In this section, we set the notation and recap the main features of the representation
of the loop-tree duality (LTD) formalism by the application of nested residues. 
Since the main idea of LTD is opening loops into connected trees, let us start the discussion
with a generic $N$-point integral at $L$ loops, in the Feynman representation, 
\begin{align}
\mathcal{A}_{N}^{\left(L\right)}\left(1,\hdots,r\right) & =\int_{\ell_{1},\hdots,\ell_{L}}d\mathcal{A}_{N}^{\left(L\right)}\left(1,\hdots,r\right)\,,
\label{eq:myamp}
\end{align}
where, to simplify the notation in the integration measure, we define 
$\int_{\ell_{s}}\equiv-\imath\mu^{4-d}\int d^{d}\ell_{s}/\left(2\pi\right)^{d}$ and, 
\begin{align}
d\mathcal{A}_{N}^{\left(L\right)}\left(1,\hdots,r\right)&=  \mathcal{N}\left(\left\{ \ell_{i}\right\} _{L},\left\{ p_{j}\right\} _{N}\right)\times G_{F}\left(1,\hdots,r\right)\,,
\label{eq:int}
\end{align}
in which, the integral is understood as an $L$-loop topology with $r$ internal lines.
The numerator $\mathcal{N}$ is a function of the loop and external momenta 
and is given by the Feynman rules of the theory. 
The function $G_F$ collects all the Feynman propagators of the topology,
\begin{align}
G_{F}\left(1,\hdots,r\right) & =\prod_{i\in1\cup2\cdots\cup r}\left(G_{F}\left(q_{i}\right)\right)^{\alpha_{i}}\,,
\end{align}
where $\alpha_{i}$ are the powers of the propagators and, 
\begin{align}
G_{F}\left(q_{i}\right) & =\frac{1}{q_{i}^{2}-m_{i}^{2}+\imath0}=\frac{1}{\left(q_{i,0}+q_{i,0}^{\left(+\right)}\right)\left(q_{i,0}-q_{i,0}^{\left(+\right)}\right)}\,,
\end{align}
the usual Feynman propagator of a one single particle, with $m_{i}$
its mass, $+\imath0$ the infinitesimal imaginary Feynman prescription
and, 
\begin{align}
q_{i,0}^{\left(+\right)} & =\sqrt{\boldsymbol{q}_{i}^{2}+m_{i}^{2}-\imath0}\,,
\end{align}
the on-shell energy of the loop momentum $q_{i}$ written in terms
of the spatial components $\boldsymbol{q}_{i}$. Let us remark that
the dependence on the energy component of the loop momenta $q_{i,0}$
is completely pulled out. This is done, in order to apply the Cauchy
residue theorem and integrate out one degree of freedom: the energy
component of each loop momentum. Hence, when taking the nested residue~\cite{Aguilera-Verdugo:2020nrp},
one selects the poles with negative imaginary part in the complex
plane of the energy component of the loop momentum that is integrated,
as extensively explained in Ref.~\cite{Verdugo:2020kzh}.

Hence, the dual representation of the integrand (\ref{eq:int}), after
setting on shell the propagators that depend on the loop momentum
$q_{i_{1}}$, is defined as follows, 
\begin{align}
\mathcal{A}_{D}^{\left(L\right)}\left(1,\hdots,r\right) & \equiv\sum_{i_{1}\in1}\text{Res}\left(d\mathcal{A}_{N}^{\left(L\right)}\left(1,\hdots,r\right),\text{Im}\left(q_{i_{1},0}\right)<0\right)\,,
\end{align}
where the factor $-2\pi\imath$ that comes from the Cauchy residue
theorem is absorbed in the definition of the integration measure as
shall be noted in the following. This residue corresponds to integrating
out the energy components of the loop momenta. Thus, we can introduce
the nested residue as, 
\begin{align}
\mathcal{A}_{D}^{\left(L\right)}\left(1,\hdots,s;s+1,\hdots,r\right) & \equiv\sum_{i_{s}\in s}\text{Res}\left(\mathcal{A}_{D}^{\left(L\right)}\left(1,\hdots,s-1;s,\hdots,r\right),\text{Im}\left(q_{i_{r},0}\right)<0\right)\,,
\label{eq:nestres}
\end{align}
where the iteration goes until the $s$-th set and corresponds to
setting simultaneously $L$ lines on shell. The latter is equivalent
to open the loop topology (or amplitude) into connected trees. 

Finally, with the integration of the energy component of the loop
momentum, one passes from Minkowski to Euclidean space. In the following,
we use the abbreviation, 
\begin{align}
\int_{\vec{\ell}_{s}}\bullet\equiv & -\mu^{d-4}\int\frac{d^{d-1}\ell_{s}}{\left(2\pi\right)^{d-1}}\bullet\,,
\end{align}
for the $\left(d-1\right)$-momentum integration measure. 

\subsection{Maximal loop topology}
The novel representation of the LTD formalism
has displayed very interesting properties at integrand level. In particular,
the causal structure of  integrands allows for a smooth numerical
evaluation and, therefore, a numerical integration. In this section,
we briefly recall the main results obtained in~\cite{Aguilera-Verdugo:2020kzc,Verdugo:2020kzh}
w.r.t. the Maximal loop topology and set the notation  
for the following sections. 

Interestingly, the LTD formalism has been observed to work at the level
of vertices, where the causal thresholds are generated, 
and is independent of the number of loops. Namely, the
structure of the integrand, after cancellations at intermediate steps,
manifest the same causal structure independently on the 
number of internal lines, provided that the number of vertices remains the same.

Let us review the simplest topology generated with 
two vertices,\footnote{
These kinds of diagrams with $L$ internal lines are usually called multi-loop sunrise
or multi-banana diagram.
}
the Maximal loop topology~(MLT), originally studied in~\cite{Verdugo:2020kzh},
whose causal representation, after applying the LTD theorem, is, 
\begin{align}
\mathcal{A}_{\text{MLT}}^{\left(L\right)} & =-\int_{\boldsymbol{\ell}_{1},\hdots,\boldsymbol{\ell}_{L}}\frac{1}{x_{L+1}}\left(\frac{1}{\lambda_{1}^{+}}+\frac{1}{\lambda_{1}^{-}}\right)\,,
\label{eq:MLT}
\end{align}
with $x_{s}=\prod_{i=1}^{s}\,2q_{i,0}^{(+)}\,$. 

In Eq.~(\ref{eq:MLT}), it is understood an $L$-loop two-point function,
with the explicit dependence of the energy of the external momentum
$p_1$ in the causal propagator $\lambda_{1}^{\pm}=q_{(1,\hdots,L+1)}^{(+)}\pm p_{1,0}$, 
with $q_{(1,\hdots,L+1),0}^{(+)}=\sum_{i=1}^{L+1}q_{i,0}^{\left(+\right)}$.
Here, and in the following, we use $\lambda_i^{\pm}$ to refer to the $i$-th causal propagator, 
with ``$\pm$'' the direction flow w.r.t. the external momenta, as discussed below.
All causal propagators are expressed as sums of $q_{i,0}^{(+)}$, 
$q_{i_1,0}^{(+)}+q_{i_2,0}^{(+)}+\hdots+q_{i_r,0}^{(+)}$,
allowing for a smooth numerical evaluation of the integrand. 
Let us remark that the application of LTD on a Feynman integrand generates 
independent terms, obtained from Eq.~\eqref{eq:nestres}, 
that are contaminated by non-causal propagators,
$q_{i_1,0}^{(+)}-q_{i_2,0}^{(+)}+\hdots-q_{i_r,0}^{(+)}$, 
and correspond to spurious singularities, whose dependence is
completely dropped in the full sum of contributions 
(see Ref.~\cite[Fig.~1]{Aguilera-Verdugo:2020kzc}).

The structure of $\lambda_{1}^{\pm}$
exactly corresponds to having all internal lines aligned in the same
direction,
\begin{align}
\mathcal{A}_{\text{MLT}}^{\left(L\right)} & = 
\parbox{20mm}{
\begin{tikzpicture}[line width=1 pt, node distance=0.8 cm and 0.4 cm]
\coordinate (v1) at (0,0);
\coordinate (v2) at (1,0);
\coordinate (v22) at (2,0);
\coordinate (pout) at (2.7,0);
\coordinate (pin) at (-0.7,0);
\coordinate (v2a) at (1,0.5);
\draw[fermionbar] (v1) -- (v22);
\semiloop[fermion]{v1}{v22}{0};
\semiloop[fermionbar]{v22}{v1}{180};
\draw[fill=black] (v1) circle (.05cm);
\draw[fill=black] (v22) circle (.05cm);
\draw[fermionbar] (pin) -- (v1);
\draw[fermion] (pout) -- (v22);
\node  at (1,1.3) {\footnotesize$1$}; 
\node  at (1,0.6) {$\huge\boldsymbol{\vdots}$}; 
\node  at (1,-0.3) {\footnotesize$L$}; 
\node  at (1,-1.3) {\footnotesize$L+1$}; 
\draw[phi] (0.5,1.3) -- (0.5,-1.3);
\end{tikzpicture}
}
\qquad\qquad+
\parbox{20mm}{
\begin{tikzpicture}[line width=1 pt,node distance=0.4 cm and 0.4 cm]
\coordinate (v1) at (0,0);
\coordinate (v2) at (1,0);
\coordinate (v22) at (2,0);
\coordinate (pout) at (2.7,0);
\coordinate (pin) at (-0.7,0);
\coordinate (v2a) at (1,0.5);
\draw[fermion] (v1) -- (v22);
\semiloop[fermionbar]{v1}{v22}{0};
\semiloop[fermion]{v22}{v1}{180};
\draw[fill=black] (v1) circle (.05cm);
\draw[fill=black] (v22) circle (.05cm);
\draw[fermion] (pin) -- (v1);
\draw[fermionbar] (pout) -- (v22);
\node  at (1,1.3) {\footnotesize$1$}; 
\node  at (1,0.6) {$\huge\boldsymbol{\vdots}$}; 
\node  at (1,-0.3) {\footnotesize$L$}; 
\node  at (1,-1.3) {\footnotesize$L+1$}; 
\draw[phi] (0.5,1.3) -- (0.5,-1.3);
\end{tikzpicture}
}
\qquad\qquad\,,
\label{eq:mltdiagram}
\end{align}
where the two possible directions of the internal lines depend on the flow of 
the external momentum. Hence, assuming that the two-point function 
is characterised by only one independent external momentum, $p_1$, 
one has two contributions, when the energy of the incoming particle is positive, $\lambda_1^{+}$,
or negative, $\lambda_1^{-}$. 
The former and the latter contributions are understood, respectively, in Eq.~\eqref{eq:MLT}. 
Notice that due to the way how $q_{i,0}^{(+)}$'s have been defined, within LTD,
they are always positive quantities. Therefore, the appearance of thresholds 
comes only from the structure of the energy of the external momenta,
\begin{align}
\lambda_1^{+}\to&\text{ threshold if } p_{1,0}>0: \quad q_{(1,\hdots,L+1),0}^{(+)} = p_{1,0}\,, \notag\\
\lambda_1^{-}\to&\text{ threshold if } p_{1,0}<0: \quad q_{(1,\hdots,L+1),0}^{(+)} = -p_{1,0}\,.
\end{align}

Let us emphasise that the representation of Eq.~\eqref{eq:MLT} is valid at 
all loop orders and it is obtained by the nested application of the Cauchy residue theorem, 
as discussed in the former section. 

To obtain expression~\eqref{eq:MLT}, we parametrise,
without loss of generality, all internal lines of an $L$-loop MLT as
 $q_i=\ell_i$, with $i=1,\hdots,L$, and $q_{L+1}=-\sum_{i=1}^{L}\ell_i-p_{1,0}$. 
Therefore, in view of this pattern that is supported by 
factorisation and convolution identities, 
we define an \textit{edge} as a set of internal lines, 
in which all elements connect two vertices, 
\begin{align}
\parbox{20mm}{
\begin{tikzpicture}[line width=1.5 pt, node distance=0.8 cm and 0.4 cm]
\coordinate (v1) at (0,0);
\coordinate (v22) at (2,0);
\draw[fermionbar] (v1) -- (v22);
\draw[fill=black] (v1) circle (.05cm);
\draw[fill=black] (v22) circle (.05cm);
\end{tikzpicture}
}
\quad \equiv \quad
\parbox{20mm}{
\begin{tikzpicture}[line width=1 pt, node distance=0.8 cm and 0.4 cm]
\coordinate (v1) at (0,0);
\coordinate (v2) at (1,0);
\coordinate (v22) at (2,0);
\coordinate (v2a) at (1,0.5);
\draw[fermionbar] (v1) -- (v22);
\semiloop[fermion]{v1}{v22}{0};
\semiloop[fermionbar]{v22}{v1}{180};
\draw[fill=black] (v1) circle (.05cm);
\draw[fill=black] (v22) circle (.05cm);
\node  at (1,1.3) {\footnotesize$1$}; 
\node  at (1,0.6) {$\huge\boldsymbol{\vdots}$}; 
\node  at (1,-0.3) {\footnotesize$s$}; 
\node  at (1,-1.3) {\footnotesize$s+1$}; 
\end{tikzpicture}
}\quad \,,
\label{eq:graphrep}
\end{align}
where the thick line in the l.h.s. corresponds to the collection of $s+1$ internal lines 
aligned in the same direction when external momenta are attached to both vertices.
We remark that, independently of the number of internal lines (set elements), 
momentum conservation has to be satisfied in each vertex.

\section{Causal representation of Feynman integrands}
\label{sec:preliminar}

In Ref.~\cite{Aguilera-Verdugo:2020kzc}, 
it was studied the causal structure of topologies 
with up to four vertices, Next-to Maximal (NMLT) and Next-to-Next-to Maximal (N$^2$MLT)
loop topologies, where a definition and identification of entangled causal propagators
were provided. 
In this section, we recap the results for the NMLT and N$^2$MLT configurations 
and establish two operations, \textit{removing} and \textit{collapsing},
to relate topologies with different number of edges and vertices.
In fact, as shall be described in the next sections, the more knowledge 
we have on an N$^k$MLT configuration, the more applications we can find
out of it. 

\subsection{Next-to Maximal and Next-to-Next-to Maximal loop topologies}

Following the notation of Eq.~\eqref{eq:graphrep}, we construct a topology made
of three vertices, 
\begin{align}
\mathcal{A}_{\text{NMLT}}^{\left(L\right)} & = \
\parbox{20mm}{
\begin{tikzpicture}
\coordinate (v1) at (0,0);
\coordinate (v2) at (0,1);
\coordinate (v3) at (0.5,0.5);
\coordinate (v4) at (-0.5,0);
\coordinate (v5) at (-0.5,1);
\coordinate (v6) at (1,0.5);
\draw[fermionnoarrow, ultra thick] (v1) -- (v2);
\draw[fermionnoarrow, ultra thick] (v1) -- (v3);
\draw[fermionnoarrow, ultra thick] (v3) -- (v2);
\draw[fill=black] (v1) circle (.05cm);
\draw[fill=black] (v2) circle (.05cm);
\draw[fill=black] (v3) circle (.05cm);
\draw[fermionbar, thick] (v1) -- (v4);
\draw[fermionbar, thick] (v2) -- (v5);
\draw[fermionbar, thick] (v3) -- (v6);
\node  at (-0.2,0.5) {\footnotesize$s_1$}; 
\node  at (0.4,0.85) {\footnotesize$s_2$}; 
\node  at (0.4,0.15) {\footnotesize$s_3$}; 
\node  at (-0.7,0) {\tiny$p_1$}; 
\node  at (-0.7,1) {\tiny$p_2$}; 
\node  at (1.2,0.5) {\tiny$p_3$}; 
\end{tikzpicture}
} 
\quad = 
\parbox{20mm}{
\begin{tikzpicture}
\coordinate (v1) at (0,0);
\coordinate (v2) at (0,1);
\coordinate (v3) at (0.5,0.5);
\coordinate (v4) at (-0.5,0);
\coordinate (v5) at (-0.5,1);
\coordinate (v6) at (1,0.5);
\draw[fermionbar, ultra thick] (v1) -- (v2);
\draw[fermionbar, ultra thick] (v1) -- (v3);
\draw[fermionbar, ultra thick] (v3) -- (v2);
\draw[fill=black] (v1) circle (.05cm);
\draw[fill=black] (v2) circle (.05cm);
\draw[fill=black] (v3) circle (.05cm);
\draw[fermion, thick] (v1) -- (v4);
\draw[fermionbar, thick] (v2) -- (v5);
\draw[fermionnoarrow, thick] (v3) -- (v6);
%
%
\draw[phi,very thick] (-0.25,0.65) -- (0.35,1);
\draw[phi,very thick] (-0.25,0.45) -- (0.35,0);
\end{tikzpicture}
} 
\!\!\!\!\!\!+
\parbox{20mm}{
\begin{tikzpicture}
\coordinate (v1) at (0,0);
\coordinate (v2) at (0,1);
\coordinate (v3) at (0.5,0.5);
\coordinate (v4) at (-0.5,0);
\coordinate (v5) at (-0.5,1);
\coordinate (v6) at (1,0.5);
\draw[fermionbar, ultra thick] (v1) -- (v2);
\draw[fermionbar, ultra thick] (v1) -- (v3);
\draw[fermion, ultra thick] (v3) -- (v2);
\draw[fill=black] (v1) circle (.05cm);
\draw[fill=black] (v2) circle (.05cm);
\draw[fill=black] (v3) circle (.05cm);
\draw[fermion, thick] (v1) -- (v4);
\draw[fermionnoarrow, thick] (v2) -- (v5);
\draw[fermionbar, thick] (v3) -- (v6);
%
%
\draw[phi,very thick] (0.35,0.15) -- (0.35,0.85);
\draw[phi,very thick] (-0.25,0.45) -- (0.35,0);
\end{tikzpicture}
} 
\!\!\!\!\!\!+
\parbox{20mm}{
\begin{tikzpicture}
\coordinate (v1) at (0,0);
\coordinate (v2) at (0,1);
\coordinate (v3) at (0.5,0.5);
\coordinate (v4) at (-0.5,0);
\coordinate (v5) at (-0.5,1);
\coordinate (v6) at (1,0.5);
\draw[fermion, ultra thick] (v1) -- (v2);
\draw[fermionbar, ultra thick] (v1) -- (v3);
\draw[fermion, ultra thick] (v3) -- (v2);
\draw[fill=black] (v1) circle (.05cm);
\draw[fill=black] (v2) circle (.05cm);
\draw[fill=black] (v3) circle (.05cm);
\draw[fermionnoarrow, thick] (v1) -- (v4);
\draw[fermion, thick] (v2) -- (v5);
\draw[fermionbar, thick] (v3) -- (v6);
%
%
\draw[phi, very thick] (-0.25,0.65) -- (0.35,1);
\draw[phi, very thick] (0.35,0.15) -- (0.35,0.85);
\end{tikzpicture}
} 
\!\!\!\!\!\!+ \text{reverse}\,,
\label{eq:picNMLT}
\end{align}
where ``reverse'' corresponds to changing the direction flow in the internal 
and external lines. 
Additionally, as in the MLT configuration~\eqref{eq:mltdiagram}, 
the red dashed lines that cut internal edges correspond to a pictorial representation
of causal propagators. 
The main difference of this topology, as well as others with more than two vertices, w.r.t. 
the MLT configuration is the presence of entangled causal thresholds.
For instance, each diagram in the r.h.s. of Eq.~\eqref{eq:picNMLT}
 contains simultaneously two causal propagators,
making its causal representation a bit more involved,
\begin{align}
\mathcal{A}_{\text{NMLT}}^{\left(L\right)}&=\int_{\boldsymbol{\ell}_{1},\hdots,\boldsymbol{\ell}_{L}}\frac{1}{x_{L+2}}\left(
\frac{1}{\lambda_{1}^{+}\lambda_{2}^{-}}
+\frac{1}{\lambda_{1}^{+}\lambda_{3}^{-}}
+\frac{1}{\lambda_{2}^{+}\lambda_{3}^{-}}
+ (\lambda_{i}^{+}\leftrightarrow\lambda_{i}^{-})
\right)\notag\\
&=\int_{\boldsymbol{\ell}_{1},\hdots,\boldsymbol{\ell}_{L}}\frac{1}{x_{L+2}}
\sum_{\substack{i,j=1\\
j\ne i
}
}^{3}\frac{1}{\lambda_{i}^{+}\lambda_{j}^{-}}\,.
\label{eq:nmlt}
\end{align}
The pictorial representation~\eqref{eq:picNMLT} or, equivalently, the expression of 
Eq.~\eqref{eq:nmlt} is obtained from the application of LTD, 
with no loss of generality, at one loop,  clearly yielding to the 
known parametric form of NMLT.
Let us remark that this approach is different from the one in~\cite{Aguilera-Verdugo:2020kzc},
where the use of the application of LTD is carried out at three loops. 
The causal propagators $\lambda_i^{\pm}$ of this topology are expressed
as follows, 
\begin{align}
 & \lambda_{1}^{\pm}=q_{\left(1,3\right),0}^{\left(+\right)}\pm p_{1,0}\,, &  & \lambda_{2}^{\pm}=q_{\left(1,2\right),0}^{\left(+\right)}\pm p_{2,0}\,, &  & \lambda_{3}^{\pm}=q_{\left(2,3\right),0}^{\left(+\right)}\pm p_{3,0}\,,
 \label{eq:lambnmlt}
\end{align}
where, as mentioned in the former section, $q_{s_{i},0}^{\left(+\right)}=\sum_{i\in s_{i}}q_{i,0}^{\left(+\right)}$,  
contains all internal lines that connects two vertices.
In effect, the causal propagators of Eq.~(3.23) of Ref.~\cite{Aguilera-Verdugo:2020kzc} 
can be extracted by considering the sets $s_1=\{1,2\},s_2=\{12\}$ and $s_3=\{3,\hdots,L+1\}$.\footnote{
We closely follow the notation of Ref.~\cite{Aguilera-Verdugo:2020kzc},
where $i\equiv\ell_i, 12\equiv -\ell_1-\ell_2+p_2$ and 
$L+1\equiv -\sum_{i=1}^{L}\ell_i-p_1$.
}
Also, we make explicit the dependence on the three external momenta, 
which are related by momentum conservation $p_1+p_2+p_3=0$. 

Let us now draw the attention to a topology built from four vertices, 
which corresponds to the N$^2$MLT configuration.
By connecting all vertices themselves, one finds,  
\begin{align}
\mathcal{A}_{\text{N}^{2}\text{MLT}}^{(L)}&=
\parbox{25mm}{
\begin{tikzpicture}
\coordinate (v1) at (0,0);
\coordinate (v2) at (0,1);
\coordinate (v3) at (1,1);
\coordinate (v4) at (1,0);
\coordinate (p1) at (-0.5,0);
\coordinate (p2) at (-0.5,1);
\coordinate (p3) at (1.5,1);
\coordinate (p4) at (1.5,0);
\draw[fermionnoarrow, ultra thick] (v1) -- (v2);
\draw[fermionnoarrow, ultra thick] (v1) -- (v4);
\draw[fermionnoarrow, ultra thick] (v2) -- (v3);
\draw[fermionnoarrow, ultra thick] (v3) -- (v4);
\draw[fermionnoarrow, ultra thick] (v1) -- (v3);
\draw[fermionnoarrow, ultra thick] (v2) -- (0.4,0.6);
\draw[fermionnoarrow, ultra thick] (v4) -- (0.6,0.4);
\draw[fill=black] (v1) circle (.05cm);
\draw[fill=black] (v2) circle (.05cm);
\draw[fill=black] (v3) circle (.05cm);
\draw[fill=black] (v4) circle (.05cm);
\draw[fermionbar, thick] (v1) -- (p1);
\draw[fermionbar, thick] (v2) -- (p2);
\draw[fermionbar, thick] (v3) -- (p3);
\draw[fermionbar, thick] (v4) -- (p4);
\node  at (-0.2,0.5) {\tiny$s_1$}; 
\node  at (0.5,1.15) {\tiny$s_2$}; 
\node  at (1.2,0.5) {\tiny$s_3$}; 
\node  at (0.5,-0.15) {\tiny$s_4$}; 
\node  at (0.4,0.2) {\tiny$s_5$}; 
\node  at (0.4,0.8) {\tiny$s_6$}; 
\node  at (-0.7,0) {\tiny$p_1$}; 
\node  at (-0.7,1) {\tiny$p_2$}; 
\node  at (1.7,1) {\tiny$p_3$}; 
\node  at (1.7,0) {\tiny$p_4$}; 
\end{tikzpicture}
} 
\qquad=
\parbox{25mm}{
\begin{tikzpicture}
\coordinate (v0) at (0.5,0.5);
\coordinate (v1) at (0.5,1);
\coordinate (v2) at (0.94,0.25);
\coordinate (v3) at (0.07,0.25);
\coordinate (p1) at (-0.4,0);
\coordinate (p2) at (0.5,1.4);
\coordinate (p3) at (1.4,0);
\coordinate (p4) at (0.5,0.2);
\draw[fermion, thick] (p1) -- (v3);
\draw[fermion, thick] (p2) -- (v1);
\draw[fermion, thick] (p3) -- (v2);
\draw[fermion, thick] (p4) -- (v0);
\draw[fill=black] (v1) circle (.05cm);
\draw[fill=black] (v2) circle (.05cm);
\draw[fill=black] (v3) circle (.05cm);
\draw[fill=black] (v0) circle (.05cm);
\draw[fermionnoarrow, ultra thick] (v0) circle (0.5);
\draw[fermionnoarrow, ultra thick] (v1) -- (v0);
\draw[fermionnoarrow, ultra thick] (v2) -- (v0);
\draw[fermionnoarrow, ultra thick] (0.07,0.25) -- (v0);
\node  at (-0.55,0) {\tiny$p_1$}; 
\node  at (0.5,1.55) {\tiny$p_2$}; 
\node  at (1.55,0) {\tiny$p_3$}; 
\node  at (0.65,0.2) {\tiny$p_4$}; 
\node  at (0.5,-0.15) {\tiny$s_5$}; 
\node  at (-0.1,0.8) {\tiny$s_1$}; 
\node  at (1.1,0.8) {\tiny$s_2$}; 
\node  at (0.8,0.45) {\tiny$s_3$}; 
\node  at (0.2,0.45) {\tiny$s_4$};
\node  at (0.65,0.75) {\tiny$s_6$};
%
\end{tikzpicture}
} 
\quad,
\label{eq:fign2mlt}
\end{align}
where the r.h.s. clearly is the ``planarised'' form of the l.h.s. and this topology 
is often called Mercedes-Benz-like diagram. 
The causal representation of the N$^2$MLT configuration is found to be, 
\begin{align}
\mathcal{A}_{\text{N}^{2}\text{MLT}}^{(L)}&=  -\int_{\ell_{1},\hdots,\ell_{L}}\frac{1}{x_{L+3}}
\left[L_{12}^{+}\,\Lambda_{34}^{-}
+L_{13}^{+}\,\Lambda_{24}^{-}
+L_{23}^{+}\,\Lambda_{14}^{-}
 +\left(\lambda^{+}_i\leftrightarrow\lambda^{-}_i\right)
 \right]\,,
 \label{eq:n2mltcausal}
\end{align}
with the shorthand notation, 
\begin{align}
\Lambda_{ij}^{\pm}=\left(\frac{1}{\lambda_{i}^{\pm}}+\frac{1}{\lambda_{j}^{\pm}}\right)\,,\qquad L_{ij}^{\pm}=\frac{1}{\lambda_{ij}^{\pm}}\left(\frac{1}{\lambda_{i}^{\pm}}+\frac{1}{\lambda_{j}^{\pm}}\right)\,.
\end{align}
We also suggestively name the causal propagators
$\lambda_{i}$ and $\lambda_{ij}$, according to their dependence on the 
external momenta, $p_i$ and $p_{ij}=p_{i}+p_{j}$, respectively, 
\begin{align}
 & \lambda_{1}^{\pm}=q_{\left(1,4,5\right),0}^{\left(+\right)}\pm p_{1,0}\,,\nonumber \\
 & \lambda_{2}^{\pm}=q_{\left(1,2,6\right),0}^{\left(+\right)}\pm p_{2,0}\,, &  & \lambda_{12}^{\pm}=q_{\left(2,4,5,6\right),0}^{\left(+\right)}\pm p_{12,0}\,,\nonumber \\
 & \lambda_{3}^{\pm}=q_{\left(2,3,5\right),0}^{\left(+\right)}\pm p_{3,0}\,, &  & \lambda_{13}^{\pm}=q_{\left(1,2,3,4\right),0}^{\left(+\right)}\pm p_{13,0}\,,\nonumber \\
 & \lambda_{4}^{\pm}=q_{\left(3,4,6\right),0}^{\left(+\right)}\pm p_{4,0}\,, &  & \lambda_{23}^{\pm}=q_{\left(1,3,5,6\right),0}^{\left(+\right)}\pm p_{23,0}
 \,,\label{eq:lambn2mlt}
\end{align}
with the momentum conservation $p_{1}+p_{2}+p_{3}+p_{4}=0$.
Allowing, in this way, to express $\mathcal{A}_{\text{N}^{2}\text{MLT}}$ in the most compact form, 
\begin{align}
\mathcal{A}_{\text{N}^{2}\text{MLT}}^{(L)}&=-\int_{\ell_{1},\hdots,\ell_{L}}\frac{1}{x_{L+3}}
 \sum_{\substack{
i=1\\
j=i+1}}^{4}
 \sum_{\substack{k=1\\
k\ne i,j
}
}^{4}L_{ij}^{+}\frac{1}{\lambda_{k}^{-}}\,.
\label{eq:n2mltcompact}
\end{align}
In order to read this sum, one has to take into account that, 
because of momentum conservation, 
\begin{align}
 \lambda_{34}^{\pm}=\lambda_{12}^{\mp}\,,
 & & \lambda_{14}^{\pm}=\lambda_{23}^{\mp}\,,
& & \lambda_{24}^{\pm}=\lambda_{13}^{\mp}\,.
\end{align}

The expression~\eqref{eq:n2mltcausal} is obtained by applying the LTD theorem on 
the three-loop Mercedes-Benz like diagram, whose internal lines are expressed 
by convenience as,
\begin{align}
\{
q_1,\hdots,q_6
\} = 
\{\ell _1,
\ell _2,-\ell _2+\ell _3-p_3,
\ell _1-\ell _3-p_1,
\ell _3,
-\ell _1+\ell _2-p_2
\}\,.
\end{align}
Let us remark that although we apply LTD on this topology with six internal lines, 
the very same behaviour, independently of the loop order, is identical when
each edge contains more than a single element. 

The notation $\lambda_i^{\pm}$ in Eq.~\eqref{eq:lambn2mlt} is slightly different to the 
one found in Ref.~\cite{Aguilera-Verdugo:2020kzc}, which is due to the way how the
internal sets $s_i$ have been constructed. 
In particular, the extension to $L$ loops in the 
latter was only carried out in the set $s_5$, by considering $s_5=\{4,\hdots,L+1\}$, 
while each of the other sets contains only one element. 
Hence, the notation,
\begin{align}
q_{\left(i_{1},i_{2},\hdots,i_{r}\right),0}^{(+)}&=\sum_{k\in i_{1}\cup i_{2}\cup\cdots\cup i_{r}}q_{k,0}^{\left(+\right)}\,,
\end{align}
is straightforwardly extended.

\subsection{Removing and collapsing procedures}

The causal representation of N$^2$MLT displays a much richer structure  than
the NMLT configuration. 
It is for this reason that the question about the existence of a relation among these loop
topologies, and extension to others with higher orders in edges and vertices,  
is well posed. Then, to understand this pattern,
we give a preliminary answer to the latter
by introducing two procedures at the level of the topologies: 
\textit{removing} and \textit{collapsing}.

\subsubsection{Removing}
\label{sec:removing}
In order to start this discussion, let us recap the way how the 
causal representation of the NMLT and N$^2$MLT configurations
have been generated. In particular, the NMLT configuration 
was obtained through the application 
of LTD on a one-loop three-point scalar integral, whereas 
the N$^2$MLT configuration was generated by the application of LTD on a
three-loop three-point scalar integral. 
However,  the very same structure of both integrands holds when
more internal lines are included and it is the reason for working 
with edges in the present manuscript. 
Notice that the only difference when considering more internal lines
(or more elements in the edges) is the presence of 
additional $q_{i,0}^{(+)}$'s in the definition of $\lambda_{i}^{\pm}$. 
Hence, understanding the removing operation of elements in an edge is straightforward:
one sets the corresponding  $q_{i,0}^{(+)}\to0$. 
Interestingly, this operation can also directly be applied on edges, 
taking into account that the resulting topology, after the removing, 
cannot be disconnected. 

Then, with this procedure, we consider the N$^2$MLT topology and remove,
with no loss of generality, the edges $s_5$ and $s_6$ (see Eq.~\eqref{eq:fign2mlt}). 
This procedure corresponds 
to setting $q_{s_5,0}^{(+)},q_{s_6,0}^{(+)}\to0$  in the definition of  $\lambda_{i}^{\pm}$'s,
\begin{align}
\mathcal{R}_{(5,6)}\left[
\parbox{22mm}{
\begin{tikzpicture}
\coordinate (v1) at (0,0);
\coordinate (v2) at (0,1);
\coordinate (v3) at (1,1);
\coordinate (v4) at (1,0);
\coordinate (p1) at (-0.5,0);
\coordinate (p2) at (-0.5,1);
\coordinate (p3) at (1.5,1);
\coordinate (p4) at (1.5,0);
\draw[fermionnoarrow, ultra thick] (v1) -- (v2);
\draw[fermionnoarrow, ultra thick] (v1) -- (v4);
\draw[fermionnoarrow, ultra thick] (v2) -- (v3);
\draw[fermionnoarrow, ultra thick] (v3) -- (v4);
\draw[fermionnoarrow, ultra thick] (v1) -- (v3);
\draw[fermionnoarrow, ultra thick] (v2) -- (0.4,0.6);
\draw[fermionnoarrow, ultra thick] (v4) -- (0.6,0.4);
\draw[fill=black] (v1) circle (.05cm);
\draw[fill=black] (v2) circle (.05cm);
\draw[fill=black] (v3) circle (.05cm);
\draw[fill=black] (v4) circle (.05cm);
\draw[fermionbar, thick] (v1) -- (p1);
\draw[fermionbar, thick] (v2) -- (p2);
\draw[fermionbar, thick] (v3) -- (p3);
\draw[fermionbar, thick] (v4) -- (p4);
\node  at (-0.2,0.5) {\tiny$s_1$}; 
\node  at (0.5,1.15) {\tiny$s_2$}; 
\node  at (1.2,0.5) {\tiny$s_3$}; 
\node  at (0.5,-0.15) {\tiny$s_4$}; 
\node  at (0.4,0.2) {\tiny$s_5$}; 
\node  at (0.4,0.8) {\tiny$s_6$}; 
%
\end{tikzpicture}
} 
\right]
\to\ \
\parbox{20mm}{
\begin{tikzpicture}
\coordinate (v1) at (0,0);
\coordinate (v2) at (0,1);
\coordinate (v3) at (1,1);
\coordinate (v4) at (1,0);
\coordinate (p1) at (-0.5,0);
\coordinate (p2) at (-0.5,1);
\coordinate (p3) at (1.5,1);
\coordinate (p4) at (1.5,0);
\draw[fermionnoarrow, ultra thick] (v1) -- (v2);
\draw[fermionnoarrow, ultra thick] (v1) -- (v4);
\draw[fermionnoarrow, ultra thick] (v2) -- (v3);
\draw[fermionnoarrow, ultra thick] (v3) -- (v4);
\draw[fill=black] (v1) circle (.05cm);
\draw[fill=black] (v2) circle (.05cm);
\draw[fill=black] (v3) circle (.05cm);
\draw[fill=black] (v4) circle (.05cm);
\draw[fermionbar, thick] (v1) -- (p1);
\draw[fermionbar, thick] (v2) -- (p2);
\draw[fermionbar, thick] (v3) -- (p3);
\draw[fermionbar, thick] (v4) -- (p4);
\node  at (-0.2,0.5) {\tiny$s_1$}; 
\node  at (0.5,1.15) {\tiny$s_2$}; 
\node  at (1.2,0.5) {\tiny$s_3$}; 
\node  at (0.5,-0.15) {\tiny$s_4$}; 
\end{tikzpicture}
} 
\quad.
\label{eq:figremoving}
\end{align}
Here and in the following, the operator $\mathcal{R}_{i_1,i_2,\hdots,i_N}[\#]$ corresponds to removing the 
edges $s_{i_1},s_{i_2},\hdots,s_{i_N}$ in the multi-loop topology.
In effect, this operation introduces simplifications in the expression~\eqref{eq:n2mltcausal},
\begin{align}
L_{13}^{\pm}\Lambda_{24}^{\mp}&\to\frac{1}{\lambda_{1}^{\pm}\lambda_{3}^{\pm}}\Lambda_{24}^{\mp}=\frac{1}{\lambda_{2}^{\mp}\lambda_{4}^{\mp}}\Lambda_{13}^{\pm}\,.
\label{eq:n2mlt21Lbox} 
\end{align}
Thus, recovering the causal representation of  a one-loop box calculated from
the standard application of LTD, with only 6 causal thresholds. 
A noteworthy feature of this operation is that the dependence on the causal threshold $\lambda_{13}^{\pm}$
is completely dropped, which is in agreement with the diagrammatic perspective, 
since no edge is connecting the external momenta $p_1$ and $p_3$ (or $p_2$ and $p_4$). 
Nonetheless, the same simplification is obtained by removing either $s_5$ or $s_6$
as is noted in the first and second term in the r.h.s. of Eq.~\eqref{eq:n2mlt21Lbox}, respectively.

In the former result, we were interested in the part of the integrand that only contains $\lambda_{i}^{\pm}$. 
However, setting any $q_{s_i,0}^{(+)}\to0$
introduces a singularity in $x_{L+3}$, 
which is healed by taking the following limit, 
\begin{align}
d\mathcal{A}_{\text{N$^2$MLT}}^{\left(L-|s_i|\right)} &= 
\lim_{q_{s_{i}}^{\left(+\right)}\to0}2q_{s_{i}}^{\left(+\right)}\,d\mathcal{A}_{\text{N$^2$MLT}}^{\left(L\right)}\,,
\label{eq:removingsingle}
\end{align}
where the integrand $d\mathcal{A}_N^{(L)}$ has previously defined in Eq.~\eqref{eq:int}
and, abusing of notation, $|s_i|$ corresponds to the number of elements of the set $s_i$.
In case of removing more internal lines, one elaborates on the application of a nested limit. 

So far we have considered topologies constructed from linear Feynman propagators, 
however, the removing procedure~\eqref{eq:removingsingle} can be extended to 
the general case with raised powers in the Feynman propagators.
For instance, removing one internal internal line, say $i$, whose propagator is raised to power $m_i$, 
one gets, 
\begin{align}
d\mathcal{A}_{\text{N\ensuremath{^{k}}MLT}}^{\left(L-1\right)} & =\frac{\left(-1\right)^{m_{i}-1}}{2^{m_{i}-1}\left(2m_{i}-3\right)!!}\lim\left(2q_{i}^{\left(+\right)}\right)^{2m_{i}-1}d\mathcal{A}_{\text{N\ensuremath{^{k}}MLT}}^{\left(L\right)}\,,
\end{align}
that nestedly can be applied to all internal lines of a given edge. 

\subsubsection{Collapsing}

In the former discussion, we provided a procedure to obtain, without the application of
the LTD formalism,  topologies that are related by the number
of vertices and regardless of the number of edges. 
It is indeed desirable to recover topologies with lower number of vertices by applying 
a similar procedure like removing. The motivation for doing this is to compute one 
complicated topology once and for all and, as a byproduct, all topologies 
with lower number of vertices. 
To this end, in this section, we introduce the \textit{collapsing} procedure that,
together with \textit{removing}, allows for a unified framework to generate
any causal representation from the richest topology with the highest number of vertices
and edges. 
Let us emphasise that the collapsing procedure is not intended to be an operation that shrinks an edge
to a point. On the contrary, every time this operation is applied, it removes one vertex and the 
edges that are attached to it. 
Thus, passing at each iteration from $n$ to $n-1$ vertices. 

In order to elaborate on the collapsing procedure, we consider the N$^2$MLT configuration and remove
the edges $s_3,s_4, s_6$ and the external momenta $p_4$ (see Eq.~\eqref{eq:fign2mlt}),
\begin{align}
\mathcal{C}_{(3,4,6)}^{(p_{4})}
\left[
\parbox{25mm}{
\begin{tikzpicture}
\coordinate (v0) at (0.5,0.5);
\coordinate (v1) at (0.5,1);
\coordinate (v2) at (0.94,0.25);
\coordinate (v3) at (0.07,0.25);
\coordinate (p1) at (-0.4,0);
\coordinate (p2) at (0.5,1.4);
\coordinate (p3) at (1.4,0);
\coordinate (p4) at (0.5,0.2);
\draw[fermion, thick] (p1) -- (v3);
\draw[fermion, thick] (p2) -- (v1);
\draw[fermion, thick] (p3) -- (v2);
\draw[fermion, thick] (p4) -- (v0);
\draw[fill=black] (v1) circle (.05cm);
\draw[fill=black] (v2) circle (.05cm);
\draw[fill=black] (v3) circle (.05cm);
\draw[fill=black] (v0) circle (.05cm);
\draw[fermionnoarrow, ultra thick] (v0) circle (0.5);
\draw[fermionnoarrow, ultra thick] (v1) -- (v0);
\draw[fermionnoarrow, ultra thick] (v2) -- (v0);
\draw[fermionnoarrow, ultra thick] (0.07,0.25) -- (v0);
\node  at (-0.55,0) {\tiny$p_1$}; 
\node  at (0.5,1.55) {\tiny$p_2$}; 
\node  at (1.55,0) {\tiny$p_3$}; 
\node  at (0.65,0.2) {\tiny$p_4$}; 
\node  at (0.5,-0.15) {\tiny$s_5$}; 
\node  at (-0.1,0.8) {\tiny$s_1$}; 
\node  at (1.1,0.8) {\tiny$s_2$}; 
\node  at (0.8,0.45) {\tiny$s_3$}; 
\node  at (0.2,0.45) {\tiny$s_4$};
\node  at (0.65,0.75) {\tiny$s_6$};
%
\end{tikzpicture}
} 
\right]
\ \to\
\parbox{25mm}{
\begin{tikzpicture}
\coordinate (v0) at (0.5,0.5);
\coordinate (v1) at (0.5,1);
\coordinate (v2) at (0.94,0.25);
\coordinate (v3) at (0.07,0.25);
\coordinate (p1) at (-0.4,0);
\coordinate (p2) at (0.5,1.4);
\coordinate (p3) at (1.4,0);
\draw[fermion, thick] (p1) -- (v3);
\draw[fermion, thick] (p2) -- (v1);
\draw[fermion, thick] (p3) -- (v2);
\draw[fill=black] (v1) circle (.05cm);
\draw[fill=black] (v2) circle (.05cm);
\draw[fill=black] (v3) circle (.05cm);
\draw[fermionnoarrow, ultra thick] (v0) circle (0.5);
\node  at (0.5,-0.15) {\tiny$s_5$}; 
\node  at (-0.1,0.8) {\tiny$s_1$}; 
\node  at (1.1,0.8) {\tiny$s_2$}; 
\end{tikzpicture}
} ,
\label{eq:figcollapsing}
\end{align}
with $\mathcal{C}_{(i_1,i_2,\hdots,i_N)}^{(p_{r_1})}[\#]$ the operator that collapses the internal
lines $i_1,i_2,\hdots,i_N$ and removes the external momenta $p_{r_1}$.
Certainly, the r.h.s. of Eq.~\eqref{eq:figcollapsing} should recover the known structure for NMLT. 
However, when removing these edges together with the external momentum, one notices
that there is a causal propagator that introduces a singularity, $\lambda_{4}^{\pm}$. 
Hence, to make a smooth transition from N$^2$MLT to NMLT, 
we firstly compute the limit when $\lambda_{4}^{\pm}\to0$ and nestedly the ones 
that involve, in this case, $s_3,s_4,s_6$ and $p_{4,0}$, 
\begin{align}
d\mathcal{A}_{\text{N}\text{MLT}} & =
-\frac{1}{2}
\lim_{s_3,s_4,s_6\to0}
\prod_{i\in s_{3}\cup s_{4}\cup s_{6} } 2q_{i,0}^{(+)}
\lim_{\lambda_{5}^{\pm}\to0} 
\lambda_{5}^{\pm}\,
d\mathcal{A}_{\text{N}^{2}\text{MLT}}\,.
\label{eq:n2mlt2nmlt}
\end{align}
The nested limits in Eq.~\eqref{eq:n2mlt2nmlt} express all $\lambda_{ij}^{\pm}$ as, 
\begin{align}
 & \lambda_{12}^{\pm}\to\lambda_{3}^{\mp}\,, &  & \lambda_{13}^{\pm}\to\lambda_{2}^{\mp}\,, &  & \lambda_{23}^{\pm}\to\lambda_{1}^{\mp}\,,
\end{align}
where the momentum conservation  $p_1+p_2+p_3=0$ is taken into account, thus,  
recovering the very same expression of Eq.~\eqref{eq:nmlt}.

Analogously, for the $\mathcal{A}_{\text{MLT}}$,  we have a nested limit, 
\begin{align}
d\mathcal{A}_{\text{MLT}} & =
\frac{1}{4}
\lim_{s_2,\hdots,s_6\to0}
\prod_{i\in s_{2}\cup \hdots\cup s_{6} } 2q_{i,0}^{(+)}
\lim_{\lambda_{3}^{\pm}\to0}
\lambda_{3}^{\pm}\,
\lim_{\lambda_{6}^{\pm}\to0} 
\lambda_{4}^{\pm}\,
d\mathcal{A}_{\text{N}^{2}\text{MLT}}\,,
\label{eq:n2mlt2mlt}
\end{align}
where $\lambda_{2}^{\mp}\to\lambda_{1}^{\pm}$ and $p_1+p_2=0$. 

\bigskip In the former discussions, we have introduced two procedures, \textit{removing} 
and \textit{collapsing}, to generate any kind of loop topology independently of the 
number of vertices and edges. 
In particular, we consider the relations between N$^2$MLT with less number of edges
and the transition from N$^2$MLT to NMLT and MLT. 
We observe that starting from an N$^k$MLT topology, with $k+2$ vertices, 
allows for a straightforward reconstruction of the causal representation 
of all lower topologies. 
It is also clear that more complexity is expected when more vertices are included.
However, as shall be noted in the following section, one can use the removing 
and collapsing procedures to understand the causal representation of N$^{k}$MLT 
from N$^{k-1}$MLT configurations. 

The most general topology with a given number of vertices is generated by considering 
all edges, namely, all possible connections among vertices, as done in NMLT and N$^2$MLT. 
The reason of being the most general as possible in the generation of topologies 
is to extract without any ambiguity the causal propagators that appear in a given topology.
The aim of the latter is to generate yet a causal representation of multi-loop topologies
without making use of the LTD formalism, by means of the application of the nested residue. 

\section{Five-vertex topologies and beyond}
\label{sec:fivetopos}

\begin{figure}[t]
\centering
\includegraphics[scale=1]{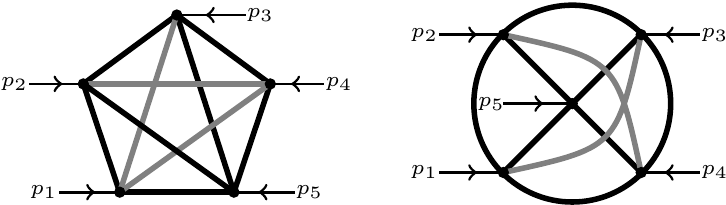}
\caption{Multi-loop topology with five vertices and ten internal lines.}
\label{eq:5cusps}
\end{figure}

Following the same discussion of the former section, we start from a topology containing 
five vertices with all possible adjacencies among themselves, 
as depicted in Fig.~\ref{eq:5cusps},
where the grey lines do not  cross among themselves. 

\subsection{Causal representation of five-vertex topologies}

In order to harness of the LTD formalism and, therefore, 
the causal representation, we need to consider a topology at six loops with 
ten denominators,
\begin{align}
q_{i}=&\begin{cases}
q_{i}=\ell_{i},\, & i=1,\hdots,6\\
\ell_{1}-\ell_{4}-\ell_{5}-p_{1}, & i=7\\
\ell_{2}-\ell_{1}-\ell_{6}-p_{2}, & i=8\\
\ell_{3}-\ell_{2}+\ell_{5}-p_{3}, & i=9\\
\ell_{4}-\ell_{3}+\ell_{6}-p_{4} & i=10
\end{cases}
\quad \,,
\end{align}
whose causal propagators are found to be, 
\begin{align}
 & \lambda_{1}^{\pm}=q_{\left(1,4,5,7\right),0}^{\left(+\right)}\pm p_{1,0}\,, &  & \lambda_{12}^{\pm}=q_{\left(2,4,5,6,7,8\right),0}^{\left(+\right)}\pm p_{12,0}\,, &  & \lambda_{24}^{\pm}=q_{\left(1,2,3,4,8,10\right),0}^{\left(+\right)}\pm p_{24,0}\,,\nonumber \\
 & \lambda_{2}^{\pm}=q_{\left(1,2,6,8\right),0}^{\left(+\right)}\pm p_{2,0}\,, &  & \lambda_{13}^{\pm}=q_{\left(1,2,3,4,7,9\right),0}^{\left(+\right)}\pm p_{13,0}\,, &  & \lambda_{35}^{\pm}=q_{\left(2,3,5,7,8,10\right),0}^{\left(+\right)}\pm p_{35,0}\,,\nonumber \\
 & \lambda_{3}^{\pm}=q_{\left(2,3,5,9\right),0}^{\left(+\right)}\pm p_{3,0}\,, &  & \lambda_{23}^{\pm}=q_{\left(1,3,5,6,8,9\right),0}^{\left(+\right)}\pm p_{23,0}\,, &  & \lambda_{34}^{\pm}=q_{\left(2,4,5,6,9,10\right),0}^{\left(+\right)}\pm p_{34,0}\,,\nonumber \\
 & \lambda_{4}^{\pm}=q_{\left(3,4,6,10\right),0}^{\left(+\right)}\pm p_{4,0}\,, &  & \lambda_{45}^{\pm}=q_{\left(3,4,6,7,8,9\right),0}^{\left(+\right)}\pm p_{45,0}\,, &  & \lambda_{25}^{\pm}=q_{\left(1,2,6,7,9,10\right),0}^{\left(+\right)}\pm p_{25,0}\,,\nonumber \\
 & \lambda_{5}^{\pm}=q_{\left(7,8,9,10\right),0}^{\left(+\right)}\pm p_{5,0}\,, &  & \lambda_{14}^{\pm}=q_{\left(1,3,5,6,7,10\right),0}^{\left(+\right)}\pm p_{14,0}\,, &  & \lambda_{15}^{\pm}=q_{\left(1,4,5,8,9,10\right),0}^{\left(+\right)}\pm p_{15,0}\,.\label{eq:lambn3mlt}
\end{align}
Let us remark that the causal representation of this diagram will contain a product of
four causal propagators, which is expected e.g.  from a one-loop pentagon
-- no internal lines in Fig.~\ref{eq:5cusps}.
Nevertheless, not all products of $\lambda_i$'s in~\eqref{eq:lambn3mlt} will appear
in the representation. In effect, there will be products of two $\lambda_{i}^{\pm}$'s, 
that cannot appear because
of the incompatible alignment of internal and external momenta. 
In other words, these two causal propagators
will correspond, from a graphical point of view, to an overlapping. 
Let us illustrate this point by considering
a possible entanglement between $\lambda_{12}^{\pm}$, $\lambda_{14}^{\pm}$
and $\lambda_{12}^{\pm}$, $\lambda_{34}^{\pm}$. 
\begin{figure}[t]
\centering
\includegraphics[scale=1]{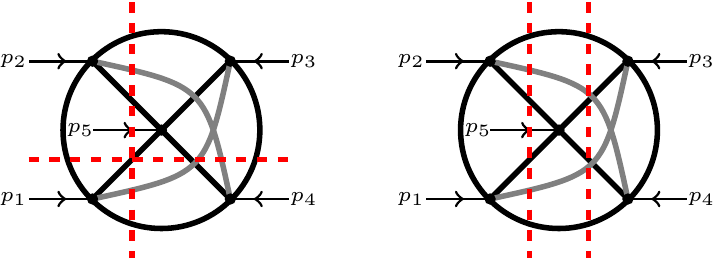}
\caption{Loop topologies with (\textit{left}:) two overlapped and 
(\textit{right}:) two entangled thresholds.}
\label{fig:overlap}
\end{figure}
In Fig.~\ref{fig:overlap} (left), we note that the causal thresholds 
$\lambda_{12}^{\pm}$, $\lambda_{14}^{\pm}$ cannot be considered together
when providing a candidate for the causal representation,
whereas, $\lambda_{12}^{\pm}$, $\lambda_{34}^{\pm}$,
in Fig.~\ref{fig:overlap} (right), appear
with combinations of other $\lambda_{i}^{\pm}$'s.
Similarly, we can apply the same procedure to tuples of length three.\footnote{
This procedure has been implemented in \Mathematica{} by studying
the direction of all causal propagators according to the presence of 
 external momenta.} 
 
Remarkably, the features described for the loop topology of Fig.~\ref{fig:overlap} 
with either overlapped or entangled causal thresholds may be related to the
Steinmann relations~\cite{Steinmann1,Steinmann2, Araki:1961, Ruelle:1961rd, Stapp:1971hh, Lassalle:1974jm, Cahill:1973qp}, since they state that double discontinuities 
in overlapping cuts vanish. Hence, from the latter there is the support
that $\lambda_{12}^{\pm}$ and $\lambda_{14}^{\pm}$
cannot appear together in the causal representation of 
this loop topology. 
The Steinmann relations have been considered in recent works~\cite{Caron-Huot:2016owq,Caron-Huot:2019bsq,Benincasa:2020aoj,Caron-Huot:2020bkp,Bourjaily:2020wvq} and 
read as follows, 
\begin{align}
&\text{Disc}_{s_{I}}\left(\text{Disc}_{s_{J}}\mathcal{A}_{N}^{\left(L\right)}\right)=0\,,\quad\text{where }
\begin{cases}
I\not\subset J\\
J\not\subset I
\end{cases}
\,,
\end{align}
where $\text{Disc}_{s_I}\mathcal{A}_{N}^{\left(L\right)}$ represents the discontinuity of 
the scattering amplitude (or Feynman integral) of Eq.~\eqref{eq:myamp} 
at the kinematic invariant 
$s_{I}=\left(\sum_{i\in I}p_{i}\right)^{2}$ generated by the external momenta
that belong to the set $I$.
A naive interpretation of the Steinmann relations within our framework
may be given by promoting the kinematic invariants $s_I$ to the 
sum of energies of external momenta present in the causal propagators.  

\bigskip
In effect, with the above considerations, one can provide an Ansatz, which, in principle,
has to be the most general one. Additionally, due to the constraints given 
by the tuples of $\lambda_{i}^{\pm}$,
one can significantly reduce the number of parameters to fit
or, equivalently, the number of linear equations to solve. 
In order to speed up their solution, we make use of the analytic reconstruction
over finite fields~\cite{vonManteuffel:2014ixa,Peraro:2016wsq,Peraro:2019svx,Klappert:2019emp}. 
In particular, with the use of the publicly  \verb"C++" code 
\FiniteFlow{}~\cite{Peraro:2019svx}.
Hence, the causal representation of the six-loop five-point scalar integral becomes, 
\begin{align}
\mathcal{A}_{\text{N}^{3}\text{MLT}}^{\left(L\right)}=&\int_{\ell_{1},\hdots,\ell_{L}}\frac{1}{x_{L+4}}
 \sum_{\substack{
i=1\\
j=i+1}}^{5}
\ \sum_{\substack{k=1\\
l=k+1\\
k,l\ne i,j
}
}^{5}
L_{ij}^{+}\,L_{kl}^{-}\,.
 \label{eq:n3mltcausal}
\end{align}
We remark that this expression has been obtained from an Ansatz and numerically checked 
with the residues generated from the nested application of LTD. 
Moreover, we can perform further checks by considering the 
removing and collapsing procedures, presented in Sec.~\ref{sec:preliminar}.

Since the expression~\eqref{eq:n3mltcausal} corresponds to N$^3$MLT,
we can begin collapsing all edges that connect the momentum $p_5$ with other vertices,
by setting to zero the edges $s_7,s_8,s_9,s_{10}$ and the external momentum $p_5$. 
This operation leads to $\lambda_{5}^{\pm}\to 0$. 
Then, looking only at the contribution proportional $(\lambda_{5}^{\pm})^{-1}$ 
in~\eqref{eq:n3mltcausal}, 
\begin{align}
\mathcal{A}_{\text{N}^{3}\text{MLT}}^{\left(L\right)}= & \int_{\ell_{1},\hdots,\ell_{L}}\frac{1}{x_{L+4}}\left(\frac{1}{\lambda_{5}^{-}}\sum_{\substack{i=1\\
j=i+1
}
}^{4}\ \sum_{\substack{k=1\\
k\ne i,j
}
}^{4}L_{ij}^{+}\,\frac{1}{\lambda_{k5}^{-}}+\left(\lambda_{i}^{+}\leftrightarrow\lambda_{j}^{+}\right)\right) 
+\mathcal{O}\left(\left(\lambda_{5}^{\pm}\right)^{0}\right)
\,,
\end{align}
and applying the collapsing procedure of Eq.~\eqref{eq:n2mlt2nmlt}, one exactly 
recovers the known result of Eq.~\eqref{eq:n2mltcausal}, 
after setting the corresponding $q_{s_i}\to0$
in the definition of the causal propagators~\eqref{eq:lambn3mlt}
or equivalently considering that $\lambda_{k5}^{\pm}\to\lambda_{k}^{\pm}$. 

\begin{table}[t]
\centering
\begin{tabular}{|r|c|c|c|}
\hline
Topology & $ \text{dim}\,\lambda_{i}^{\pm}$ & Entanglement & \#~Causal terms \\
\hline\hline
\parbox{20mm}{\includegraphics[scale=0.5]{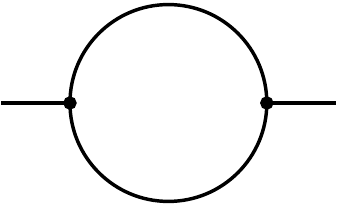}} & 1 & 1 & 2\\
\hline
\parbox{17mm}{\includegraphics[scale=1]{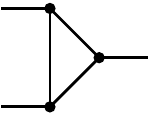}} & 3 & 2 & 6\\
\hline
\parbox{20mm}{\includegraphics[scale=1]{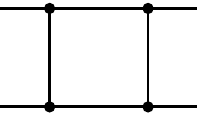}} & 6 & 3 & 20\\
\hline
\parbox{20mm}{\includegraphics[scale=0.5]{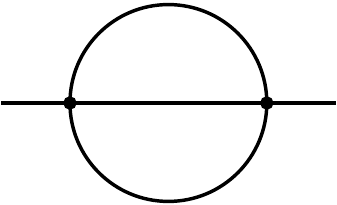}} & 1 & 1 & 2\\
\hline
\parbox{20mm}{\includegraphics[scale=1]{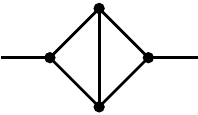}} & 6 & 3 & 20\\
\hline
\parbox{20mm}{\includegraphics[scale=1]{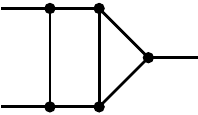}} & 10 & 3 & 70\\
\hline
\parbox{20mm}{\includegraphics[scale=1]{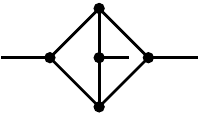}} & 11 & 3 & 76\\
\hline
\parbox{20mm}{\includegraphics[scale=1]{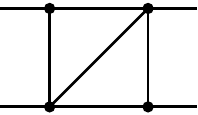}} & 6 & 3 & 20\\
\hline
\parbox{20mm}{\includegraphics[scale=1]{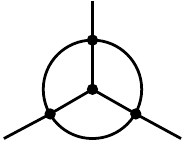}} & 7 & 3 &24\\
\hline
\parbox{20mm}{\includegraphics[scale=0.6]{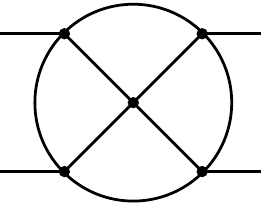}} & 13 & 4 & 98\\
\hline
\end{tabular}
\caption{Selected topologies generated from the causal representation 
of the N$^3$MLT configuration by
applying the \textit{removing} and \textit{collapsing} procedures.
}
\label{table:mytopos}
\end{table}

\bigskip In Ref.~\cite{Ramirez-Uribe:2020hes}, it has been recently computed 
the multi-loop LTD decomposition of a scattering amplitude at four loops. 
On top of it, a causal representation of the four-loop topology, with five external momenta
and eight internal lines,
was also presented, which is expressed in terms of thirteen causal propagators. 
In fact, the topology we are referring to corresponds to the diagram of Fig~\ref{eq:5cusps}
without grey lines.
Furthermore, within the \textit{removing} procedure presented in Sec.~\ref{sec:preliminar},
we obtain the causal representation by simply setting to zero 
the lines $q_{s_5}$ and $q_{s_6}$. 
By doing so, and looking at the definition of causal propagators~\eqref{eq:lambn3mlt},
we note that, in principle, we get more $\lambda_{i}^{\pm}$ than expected. 
Also, the expression in Eq.~\eqref{eq:n3mltcausal} contains 120 terms instead of 98,
as in the four-loop diagram.
Nonetheless, the same pattern, already seen in Eq.~\eqref{eq:n2mlt21Lbox} 
for the transition of a three-loop Mercedes-Benz like diagram to 
a one-loop box, is found here. In particular,
\begin{align}
 & L_{13}^{\pm}\to\frac{1}{\lambda_{1}^{\pm}\lambda_{3}^{\pm}}\,,
 && L_{24}^{\pm}\to\frac{1}{\lambda_{2}^{\pm}\lambda_{4}^{\pm}}\,,
\end{align}
dropping the dependence, in the causal representation, on $\lambda_{13}^{\pm}$
and $\lambda_{24}^{\pm}$, and, thus, recovering the 98 terms.\footnote{
Let us also comment that for the topology with five vertices and eight edges,
we have experimented an alternative approach based on the 
partial fractioning through the automated codes 
\textsc{MultivariateApart}~\cite{Heller:2021qkz}
and the \verb"Singular"~\cite{DGPS} library \verb"pfd.lib"~\cite{Boehm:2020ijp},
finding 102 causal terms instead of 98. 
}

An alternative explanation to this cancellation, directly from the diagram of Fig.~\ref{eq:5cusps}, 
can be given 
by noticing that the external momenta $p_1,p_3$ or, equivalently, $p_2,p_4$
are not connected among themselves. 
Hence, no threshold singularity may be generated from these external momenta. 

Therefore, based on the above-mentioned considerations, we remark that the 
application of a four-loop five-point scalar diagram, with eight internal lines, 
studied in Ref.~\cite{Ramirez-Uribe:2020hes}
is a particular case of the $\mathcal{A}_{\text{N}^3\text{MLT}}^{\left(L\right)} $ 
presented in this section. 
We  also observe that the former cannot reconstruct the full structure 
of $\mathcal{A}_{\text{N}^2\text{MLT}}^{\left(L\right)} $,
by using the collapsing procedure.
It, however, can reconstruct the causal representation of a one-loop box,
by removing all internal lines that connect to e.g. $p_5$.
On top of it, by applying the removing and collapsing procedures
on $\mathcal{A}_{\text{N}^3\text{MLT}}^{\left(L\right)} $,  
we generate the causal representation of topologies of up-to four loops.
These topologies are summarised in Table~\ref{table:mytopos},
where $ \text{dim}\,\lambda_{i}^{\pm}$, as shall be explained in the following section,  
corresponds to the number of causal propagators present
in a given topology. Entanglement accounts for the number of causal propagators present in each 
causal term.
\#~Causal terms corresponds to the number of terms expressed in terms of causal propagators. 
For instance, the one-loop box discussed in Sec.~\ref{sec:removing} has 
$ \text{dim}\,\lambda_{i}^{\pm}=6$ (causal thresholds),
$\text{Entanglenment}=3$
and $\text{\# Causal terms} = 20$.

\subsection{Causal propagators beyond five vertices}
In the former sections, we study the causal representation of 
$\mathcal{A}_{\text{N}^k\text{MLT}}^{\left(L\right)} $, with $k=0,1,2,3$, 
finding an interesting pattern in the structure of $\lambda_{i}^{\pm}$. 
In the following, we summarise our findings
and generalise to the $n$-vertex case.
\begin{enumerate}[label=(\roman*)]
\item \label{con:1}
A topology with $n$ vertices, containing $n$ external momenta, has 
\begin{align}
\text{dim}\,\lambda_{i}^{\pm}= &2^{n-1}-1\,,
\label{eq:Nlamb}
\end{align}
with $\text{dim}\,\lambda_{i}^{\pm}$ the number of causal propagators presented in 
each topology. 
In the results of Eqs.~\eqref{eq:lambnmlt},~\eqref{eq:lambn2mlt} and~\eqref{eq:lambn3mlt},
we use momentum conservation to have the first $n$ causal propagators in terms
of a single external momentum. Namely, that the largest subset of external momenta, 
$\{p_1,p_2,\hdots,p_{n-1}\}$, 
is replaced by a subset with only one element,~$\{p_{n}\}$.

Let us remark that the numbers in Eq.~\eqref{eq:Nlamb} accounts for the
causal propagators when all possible connections among vertices are taken into account. 
Hence, if there are tuples of vertices that are not 
connected, they have to be subtracted out from the initial counting. 
This is indeed the case of the five-vertex diagram of Fig~\ref{eq:5cusps}, 
with and without the grey edges, since in the latter the external 
momenta $p_2,p_4$ and $p_1,p_3$ are not connected. 
Therefore, these subsets have to be removed from~\eqref{eq:Nlamb},
as found in the previous section by considering the direct application of LTD.

\item\label{con:2}
The structure of each causal propagator is straightforwardly obtained, 
provided that all subsets of external momenta are properly generated. 
As mention in~\ref{con:1}, we make use of momentum conservation 
to have subsets of momenta with the minimal number of elements. 
Thus, for a topology with $n$ external momenta, 
we consider subsets with up to $[n/2]$\footnote{$[x]$ corresponds to the integral part
or integer part of $x$.} 
elements, 
allowing for a straightforward generation of all causal propagators. 

For subsets with only one element, e.g. $p_i$, we only 
need to consider the internal lines that are connecting the momentum $p_{i}$.
Similarly, for subsets with more than one element, 
say $p_i$ and $p_j$, we collapse these momenta in a single one, $p_{ij}$, 
and consider all the internal lines that are connecting to this ``new'' external momentum,
taking into account that the connecting line between $p_i$ and $p_j$
is not considered. 
\end{enumerate}

\begin{table}[t]
\centering
\begin{tabular}{|c|c|c|c|c|}
\hline & Vertices & Edges & Loops & $\text{dim}\,\lambda_{i}^{\pm}$ \\
\hline\hline
\parbox{11mm}{\includegraphics[scale=1]{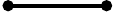}} & $2$ & $1$ & $0$ & $1$\\
\hline
\parbox{6mm}{\includegraphics[scale=1]{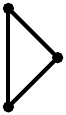}} & $3$ & $3$ & $1$ & $3$\\
\hline
\parbox{10mm}{\includegraphics[scale=1]{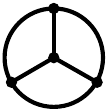}} & $4$ & $6$ & $3$ & $7$\\
\hline
\parbox{10mm}{\includegraphics[scale=0.5]{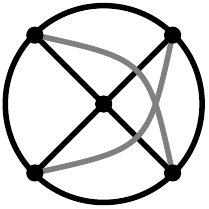}} & $5$ & $10$ & $6$ & $15$\\
\hline
$\vdots$ & $\vdots$ & $\vdots$ & $\vdots$ & $\vdots$ \\
\hline
\parbox{12mm}{\includegraphics[scale=0.5]{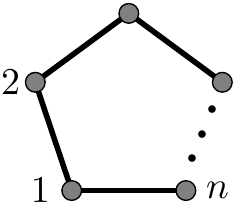}} & $n$ & $\binom{n}{2}$ & $\frac{(n-1)(n-2)}{2}$ & $2^{n-1}-1$\\
\hline
\end{tabular}
\caption{Features of topologies with Maximal number of edges, loops and causal propagators
according to the number of vertices. 
In the topology with $n$ vertices, the grey vertices account for all possible connections among
vertices.}
\label{table:chartopos}
\end{table}

\noindent Notably, from~\ref{con:1}, we can know all features of the most 
general topology made of $n$ vertices, because the maximum number of edges, loops and 
causal propagators are directly related to the number of vertices, as it is summarised
in Table~\ref{table:chartopos} for all topologies studied until now
as well as the generalisation to $n$ vertices. 

In order to illustrate these points, let us consider, for illustrative reasons,
a two-loop planar double box with six external legs.
\begin{figure}[t]
\centering
\includegraphics[scale=1.2]{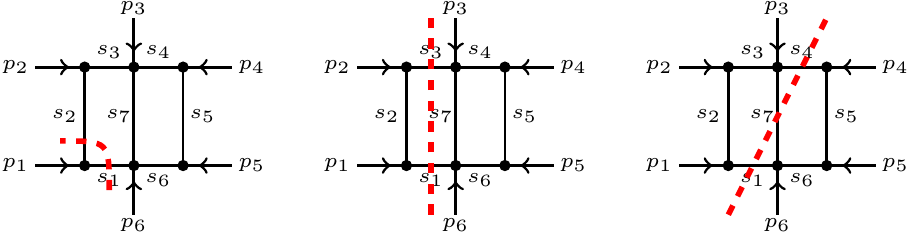}
\caption{Particular two-loop topology with six vertices and seven internal lines.}
\label{fig:2Lbox}
\end{figure}
Hence, according to~\ref{con:1}, one starts with 31 causal propagators, nevertheless, 
only 15 true ones correspond to the possible threshold singularities that can be 
obtained in a topology of seven propagators.
Thus, taking into account~\ref{con:2}, we find, 
\begin{align}
 & \lambda_{1}^{\pm}=q_{\left(1,2\right)}^{\left(+\right)}\pm p_{1,0}\,, &  & \lambda_{6}^{\pm}=q_{\left(1,6,7\right)}^{\left(+\right)}\pm p_{6,0}\,, &  & \lambda_{45}^{\pm}=q_{\left(4,6\right)}^{\left(+\right)}\pm p_{45,0}\,,\nonumber \\
 & \lambda_{2}^{\pm}=q_{\left(2,3\right)}^{\left(+\right)}\pm p_{2,0}\,, &  & \lambda_{12}^{\pm}=q_{\left(1,3\right)}^{\left(+\right)}\pm p_{12,0}\,, &  & \lambda_{56}^{\pm}=q_{\left(1,5,7\right)}^{\left(+\right)}\pm p_{56,0}\,,\nonumber \\
 & \lambda_{3}^{\pm}=q_{\left(3,4,7\right)}^{\left(+\right)}\pm p_{3,0}\,, &  & \lambda_{16}^{\pm}=q_{\left(2,6,7\right)}^{\left(+\right)}\pm p_{16,0}\,, &  & \lambda_{123}^{\pm}=q_{\left(1,4,7\right)}^{\left(+\right)}\pm p_{123,0}\,,\nonumber \\
 & \lambda_{4}^{\pm}=q_{\left(4,5\right)}^{\left(+\right)}\pm p_{4,0}\,, &  & \lambda_{23}^{\pm}=q_{\left(2,4,7\right)}^{\left(+\right)}\pm p_{23,0}\,, &  & \lambda_{234}^{\pm}=q_{\left(2,5,7\right)}^{\left(+\right)}\pm p_{234,0}\,,\nonumber \\
 & \lambda_{5}^{\pm}=q_{\left(5,6\right)}^{\left(+\right)}\pm p_{5,0}\,, &  & \lambda_{34}^{\pm}=q_{\left(3,5,7\right)}^{\left(+\right)}\pm p_{34,0}\,, &  & \lambda_{345}^{\pm}=q_{\left(3,6,7\right)}^{\left(+\right)}\pm p_{345,0}\,.\label{eq:lamb2Lbox}
\end{align}
The structure of all $\lambda_{i}^{\pm}$ in Eq.~\eqref{eq:lamb2Lbox}
are in agreement with the ones found with 
the standard application of the nested residue implemented 
in \Lotty~\cite{TorresBobadilla:2021dkq}.

\section{Causal representation of loop topologies without LTD}
\label{sec:noltd}

In Secs.~\ref{sec:preliminar} and~\ref{sec:fivetopos}, 
we study the causal representation of loop topologies
made of up-to five vertices, by following the application of LTD. Interestingly,
the structure of these topologies, thanks to the knowledge of all
causal propagators, can be inferred by properly accommodating tuples
of $\lambda_{i}^{\pm}$. This is carried out by following the approaches
summarised in Fig.~\ref{fig:overlap} and the items~\ref{con:1} and~\ref{con:2}. 
In this section, we
generate the causal representation of loop topologies made of up-to
nine vertices, without making use of LTD, but considering only the collapsing
procedure. The latter is to cross-check the result, obtained from
an Ansatz, of an N$^{k}$MLT with the corresponding lower topology,
N$^{k-1}$MLT. 

Since the relevant information of the causal representation of a given
topology only comes from the part that contains the causal propagators,
$\lambda_{i}^{\pm}$, we do not consider the prefactors $x_{L+k}$.
The latter indeed can be recovered at the end of the calculation. 
\begin{itemize}
\item The MLT configuration can only have a causal threshold. Hence, 
\begin{align}
d\mathcal{A}_{\text{MLT}}^{\left(L\right)}\sim & \frac{1}{\lambda_{1}^{+}}+\frac{1}{\lambda_{1}^{-}}\,,
\end{align}
\item The NMLT configuration has 3 causal propagators that do not overlap
among themselves. The only subtlety that one has to account for is
the direction of $\lambda_{i}^{\pm}$. Then, by organising these causal
propagators in couples, 
\begin{align}
d\mathcal{A}_{\text{NMLT}}^{\left(L\right)}\sim & \sum_{\sigma_{ij}}\frac{1}{\lambda_{i}^{+}\lambda_{j}^{-}}\,,
\end{align}
with $\sigma_{ij}=\left\{ \{1,2\},\{1,3\},\{2,1\},\{2,3\},\{3,1\},\{3,2\}\right\} $. 
\item The N$^{2}$MLT configuration has 7 causal propagators that working
with subsets can be understood as, 
\begin{align}
\vec{\lambda}_{i}^{\pm}= & \{\{1\},\{2\},\{3\},\{4\},\{1,2\},\{1,3\},\{2,3\}\}\,,
\end{align}
where, abusing of notation, $\vec{\lambda}_{i}^{\pm}$ is an array
with all causal propagators. Let us emphasise that we are interested
in the structure of the latter given by the external momenta. For
this reason, the expression of the edges 
in terms of $q_{s_i,0}^{(+)}$ 
is not needed. Noticed that two
causal propagators that have two external momenta, e.g. $\{1,2\}$
and $\{1,3\}$, cannot be entangled. Thus, we have, 
\begin{align}
d\mathcal{A}_{\text{N\ensuremath{^{2}}MLT}}^{\left(L\right)}\sim & \sum_{\substack{i=1\\
j=i+1
}
}^{4}\frac{1}{\lambda_{ij}^{+}}f_{ij}\left(\left\{ \lambda^{\pm}\right\} \right)\,,
\end{align}
where in the test function $f_{ij}$, we are left with causal propagators
with a single external momentum, $\{i\}$. Then, thanks to this notation,
it is clear that $\lambda_{i}^{\pm}$ (or $\lambda_{j}^{\pm}$) can
be ``entangled'' with $\lambda_{ij}^{\pm}$, thus, 
\begin{align}
d\mathcal{A}_{\text{N\ensuremath{^{2}}MLT}}^{\left(L\right)}\sim & \sum_{\substack{i=1\\
j=i+1
}
}^{4}\frac{1}{\lambda_{ij}^{+}}\left(\frac{1}{\lambda_{i}^{+}}+\frac{1}{\lambda_{j}^{+}}\right)\bar{f}_{ij}\left(\left\{ \lambda_{k}^{\pm}\right\} \right)=\sum_{\substack{i=1\\
j=i+1
}
}^{4}L_{ij,i,j}^{+}\bar{f}_{ij}\left(\left\{ \lambda_{k}^{\pm}\right\} \right)\,,
\end{align}
but also to causal propagators whose flow is in the opposite direction,
e.g., 
\begin{align}
 & \frac{1}{\lambda_{12}^{+}}\left(\frac{1}{\lambda_{1}^{+}}+\frac{1}{\lambda_{2}^{+}}\right)\left(\frac{1}{\lambda_{3}^{-}}+\frac{1}{\lambda_{4}^{-}}\right)\,.
\end{align}
This consideration amounts to the results found in Eq.~\eqref{eq:n2mltcompact}, 
\begin{align}
d\mathcal{A}_{\text{N\ensuremath{^{2}}MLT}}^{\left(L\right)}\sim & \sum_{\substack{i=1\\
j=i+1
}
}^{4}\sum^{4}_{\substack{k=1\\
k\ne i,j
}
}L_{ij}^{+}\frac{1}{\lambda_{k}^{-}}\,.
\end{align}
\item Let us now draw our attention to the N$^{3}$MLT, whose 15 causal
propagators can be chosen to be, 
\begin{align}
\vec{\lambda}_{i}^{\pm}= & \big\{\{1\},\{2\},\{3\},\{4\},\{5\},\nonumber \\
 & \{1,2\},\{1,3\},\{1,4\},\{1,5\},\{2,3\},\{2,4\},\{2,5\},\{3,4\},\{3,5\},\{4,5\}\big\}\,.
\end{align}
Differently to the N$^{2}$MLT configuration, there will be products
of pairs $\lambda_{ij}^{\pm}$ that will be allowed. In fact, 
\begin{align}
 & \big\{\{\{1,2\},\{3,4\}\},\{\{1,2\},\{3,5\}\},\{\{1,2\},\{4,5\}\},\{\{1,3\},\{2,4\}\},\{\{1,3\},\{2,5\}\},\nonumber \\
 & \{\{1,3\},\{4,5\}\},\{\{1,4\},\{2,3\}\},\{\{1,4\},\{2,5\}\},\{\{1,4\},\{3,5\}\},\{\{1,5\},\{2,3\}\},\nonumber \\
 & \{\{1,5\},\{2,4\}\},\{\{1,5\},\{3,4\}\},\{\{2,3\},\{4,5\}\},\{\{2,4\},\{3,5\}\},\{\{2,5\},\{3,4\}\}\big\}\,,
 \label{eq:pairn2mlt}
\end{align}
where, due to the knowledge of the N$^{2}$MLT, each subset can be
understood as, 
\begin{align}
\{\{i,j\},\{k,l\}\} & \to L_{ij}^{+}L_{kl}^{-}+L_{ij}^{-}L_{kl}^{+}\,.
\end{align}
Thus, recovering the results of Eq.~\eqref{eq:n3mltcausal},
\begin{align}
d\mathcal{A}_{\text{N}^{3}\text{MLT}}^{\left(L\right)}\sim&
 \sum_{\substack{
i=1\\
j=i+1}}^{5}
\ \sum_{\substack{k=1\\
l=k+1\\
k,l\ne i,j
}
}^{5}
L_{ij}^{+}L_{kl}^{-}\,.
\end{align}

\item The results previously described were computed by means
of the application of LTD on topologies of up-to six loops, then the cross-check
is straightforward. Let us now consider a more involved topology, the
N$^{4}$MLT, whose 31 causal propagators can be chosen to be, 
\begin{align*}
\vec{\lambda}_{i}^{\pm}= & \big\{\{1\},\{2\},\{3\},\{4\},\{5\},\{6\},\\
 & \{1,2\},\{1,3\},\{1,4\},\{1,5\},\{1,6\},\{2,3\},\{2,4\},\{2,5\},\{2,6\},\{3,4\},\{3,5\},\\
 & \{3,6\},\{4,5\},\{4,6\},\{5,6\},\{1,2,3\},\{1,2,4\},\{1,2,5\},\{1,3,4\},\{1,3,5\},\\
 & \{1,4,5\},\{2,3,4\},\{2,3,5\},\{2,4,5\},\{3,4,5\}\big\}\,.
\end{align*}
Firstly, by studying tuples of three elements, we note that none of them
can be entangled together, then, 
\begin{align}
d\mathcal{A}_{\text{N\ensuremath{^{4}}MLT}}^{\left(L\right)} & \sim\sum_{\substack{i=1\\
j=i+1\\
k=j+1
}
}^{6}\frac{1}{\lambda_{ijk}^{+}}f_{ijk}\left(\left\{ \lambda_{rs}\right\} ,\left\{ \lambda_{r}\right\} \right)\,,\label{n4mlt3}
\end{align}
Now considering the product of pairs, similar to (\ref{eq:pairn2mlt}),
we find 45 subsets of the form,
\begin{align}
 & \big\{\{\{1,2\},\{3,4\}\},\{\{1,2\},\{3,5\}\},\{\{1,2\},\{3,6\}\},\{\{1,2\},\{4,5\}\},\{\{1,2\},\{4,6\}\},\nonumber \\
 & \{\{1,2\},\{5,6\}\},\{\{1,3\},\{2,4\}\},\{\{1,3\},\{2,5\}\},\{\{1,3\},\{2,6\}\},\{\{1,3\},\{4,5\}\},\nonumber \\
 & \{\{1,3\},\{4,6\}\},\{\{1,3\},\{5,6\}\},\{\{1,4\},\{2,3\}\},\{\{1,4\},\{2,5\}\},\{\{1,4\},\{2,6\}\},\nonumber \\
 & \hdots\big\}\,.
 \label{eq:n4mlt2}
\end{align}
Notice that the intersection of the elements of the subset $\{\{i,j\},\{k,l\}\}$
is empty, $\{i,j\}\cap\{k,l\}=\emptyset$. This operation has been
implemented in \Mathematica{} for any number of vertices.

Finally, we combine (\ref{n4mlt3}) and (\ref{eq:n4mlt2}), bearing
in mind that for three given subsets, $\{i,j,k\},\{r,s\},\{t,u\}$,
one has to satify,
\begin{align}
 & \begin{array}{c}
\{r,s\}\subseteq\{i,j,k\}\\
\{t,u\}\cap\{i,j,k\}=\emptyset
\end{array}
\ .
\label{eq:setn3mlt}
\end{align}
Then, the N$^{4}$MLT configuration can be expressed as follows, 
\begin{align}
d\mathcal{A}_{\text{N\ensuremath{^{4}}MLT}}^{\left(L\right)} & \sim\sum_{\substack{i=1\\
j=i+1\\
k=j+1
}
}^{6}
\
\sum_{\substack{t=1\\
u=t+1\\
u,t\ne i,j,k
}
}^{6}L_{ijk}^{+}L_{tu}^{-}\,,
\end{align}
with, 
\begin{align}
L_{ijk}^{\pm}=\frac{1}{\lambda_{ijk}^{\pm}}\left(L_{ij}^{\pm}+L_{ik}^{\pm}+L_{jk}^{\pm}\right)\,.
\end{align}
We remark that, in order to check the validity of this relation, we have used the 
collapsing procedure to recover the expression of Eq.~\eqref{eq:setn3mlt}.
Likewise, we use the removing procedure to generate the causal representation 
of the $s$-, $t$- and $u$-like diagrams of Ref.~\cite{Ramirez-Uribe:2020hes}.

\item Similarly, for N$^{5}$MLT,  N$^{6}$MLT, and N$^{7}$MLT,
\begin{subequations}
\begin{align}
d\mathcal{A}_{\text{N\ensuremath{^{5}}MLT}}^{\left(L\right)}\sim & \sum_{\substack{i_{1}<i_{2}<i_{3}\\
j_{1}<j_{2}<j_{3}
}
}^{7}\Omega_{i_{1}i_{2}i_{3}}^{j_{1}j_{2}j_{3}}\,L_{i_{1}i_{2}i_{3}}^{+}L_{j_{1}j_{2}j_{3}}^{-}\,,\\
d\mathcal{A}_{\text{N\ensuremath{^{6}}MLT}}^{\left(L\right)}\sim & \sum_{\substack{i_{1}<i_{2}<i_{3}<i_{4}\\
j_{1}<j_{2}<j_{3}
}
}^{8}\Omega_{i_{1}i_{2}i_{3}i_{4}}^{j_{1}j_{2}j_{3}}\,L_{i_{1}i_{2}i_{3}i_{4}}^{+}L_{j_{1}j_{2}j_{3}}^{-}\,,\\
d\mathcal{A}_{\text{N\ensuremath{^{7}}MLT}}^{\left(L\right)}\sim & \sum_{\substack{i_{1}<i_{2}<i_{3}<i_{4}\\
j_{1}<j_{2}<j_{3}<j_{4}
}
}^{9}\Omega_{i_{1}i_{2}i_{3}i_{4}}^{j_{1}j_{2}j_{3}j_{4}}\,L_{i_{1}i_{2}i_{3}i_{4}}^{+}L_{j_{1}j_{2}j_{3}j_{4}}^{-}\,,
\end{align}
\label{eq:n5mtl2n7mlt}
\end{subequations}
where, in order to simplify the notation in the indices of the sum, we introduce 
a function, $\Omega$,  that can be either $0$ or $1$, depending on the intersection 
of the sets, $\vec{i}=\{i_1,i_2,\hdots,i_{N_i}\}$ and $\vec{j}=\{j_1,j_2,\hdots,j_{N_j}\}$,
\begin{align}
\Omega_{\vec{i}}^{\vec{j}}&=
\begin{cases}
1 & \text{If }\vec{i}\cap\vec{j}=\emptyset\\
0 & \text{otherwise}
\end{cases}\,,
\end{align}
and,
\begin{align}
L_{i_{1}i_{2}i_{3}i_{4}}^{\pm}&=\frac{1}{\lambda_{i_{1}i_{2}i_{3}i_{4}}^{\pm}}\left(L_{i_{1}i_{3}i_{4}}^{\pm}+L_{i_{1}i_{2}i_{4}}^{\pm}+L_{i_{1}i_{2}i_{3}}^{\pm}+L_{i_{2}i_{3}i_{4}}^{\pm}\right)\,.
\end{align}
\end{itemize}

Hence, with the above considerations, the extension 
of the causal representation of a loop topology built from $k+2$ vertices and connecting 
all internal lines, N$^{k}$MLT, can be conjectured to be, 
\begin{align}
d\mathcal{A}_{\text{N\ensuremath{^{k}}MLT}}^{\left(L\right)}\sim & \sum_{\substack{i_{1}\ll i_{N_{i}}\\
j_{1}\ll j_{N_{j}}
}
}^{k+2}\Omega_{\vec{i}}^{\vec{j}}\ L_{i_{1}i_{2}\hdots i_{N_{i}}}^{+}L_{j_{1}j_{2}\hdots j_{N_{j}}}^{-}\,,
\end{align}
where $i_{1}\ll i_{N_{i}}$ is the lexicographic ordering $i_{1}<i_{2}<\cdots<i_{N_{i}}$
and $N_{i}=\left[k/2\right]+1$ and $N_{j}=k-\left[k/2\right]$. 

\bigskip
In Eqs.~\eqref{eq:n5mtl2n7mlt}, we provide the explicit causal representation for 
N$^4$MLT, N$^5$MLT,
N$^6$MLT and N$^7$MLT without applying the LTD formalism. 
In fact, applying LTD on these topologies turns out to be not efficient
and an unnecessary calculation, which is due to the large number of loop momenta.
Let us recall that to compute the most general causal representation of these topologies, 
one needs to consider, as summarised in Table~\ref{table:chartopos}, 
10, 15, 21 and  28 loop momenta, respectively.
However, to cross-check our expressions, we make use of the removing 
procedure presented in Sec.~\ref{sec:preliminar} and compare our 
results with the ones expected from the nested application of the residue
at lower loops.
To do so, we consider the following loop topologies: 
\begin{itemize}
\item one-loop topologies with up-to nine external momenta, 
\item two-loop topologies with up-to seven external momenta,
\item ladder triangles up-to four loops,
\item three-loop tennis-court like diagram with four external momenta,
\item four-loop fishnet with four external momenta.
\end{itemize}
In order to make the comparison of the general results presented in this section with
the particular cases obtained by applying LTD, we focus only on the numerical value 
of the integrand, finding completely agreement.
Additionally, to render the integrand numerically stable, we consider the energy component
of the external momenta being imaginary. Thus, no singularities appear when evaluating
the results obtained from LTD, which at first corresponds to individual residues containing
causal and non-causal propagators, as carefully explained in Ref.~\cite{Aguilera-Verdugo:2020kzc}. 

\section{Conclusions}
\label{sec:conclusiones}

The novel representation of the loop-tree duality (LTD) formalism that started in Ref.~\cite{Verdugo:2020kzh},
has allowed for an understanding of an integrand representation displaying only 
physical information, the so-called causal representation of multi-loop Feynman integrands,
introduced in Ref.~\cite{Aguilera-Verdugo:2020kzc} for three master topologies:
Maximal (MLT), Next-to Maximal (NMLT) and Next-to-Next-to Maximal (N$^2$MLT) loop topologies. 
In particular, the behaviour at an arbitrary loop order was considered, yet obtaining the same 
representation in terms of causal propagators. 
In this paper, we performed a classification of these topologies by introducing the notion
of the features a loop topology is characterised by: \textit{vertices} and \textit{edges}. 
While the former can be related at most as the maximum number of external legs, 
the latter accounts for the collection of all internal lines that connect two vertices. 
In effect, in this paper, we recalled the causal representation of 
MLT, NMLT and N$^2$MLT in terms of vertices and edges.
We found that the results of Ref.~\cite{Aguilera-Verdugo:2020kzc} 
can be straightforwardly be obtained 
from the general form of our expressions. 

In view of the results of the MLT, NMLT and N$^2$MLT configurations, with the presence
of external momenta, we investigated the relations that may exit among themselves.
We noted that from N$^2$MLT, being the most general topology of them, we could obtain,
without performing the nested application of the LTD formalism~\cite{Aguilera-Verdugo:2020nrp},
the causal representation of  topologies with less edges and/or less vertices. 
To this end, we defined two procedures: \textit{removing} and \textit{collapsing}.
In the \textit{removing} procedure, one is interested in only removing edges
by keeping fixed the number of vertices. For instance, a one-loop box was
obtained by removing two edges in the causal representation of N$^2$MLT. 
Likewise, the \textit{collapsing} procedure reduces the number of vertices,
allowing, in this way, to obtain NMLT and MLT from N$^2$MLT. 

Furthermore, we considered the most general topologies made of $n$ vertices,
in which the number of edges accounts for all possible connections among vertices.
Within the approaches presented in this paper, we extended the results 
of the two four-loop topologies with eight and nine propagators presented in Ref.~\cite{Ramirez-Uribe:2020hes}. 
In fact, these topologies correspond to particular cases of our 
N$^3$MLT and N$^4$MLT configurations, where the \textit{removing} procedure was utilised.
Thus, to check our results, we used the \textit{collapsing} and \textit{removing}
procedures to find a topology with lower complexity.
In other words, by computing the most general N$^k$MLT configuration with $k+2$ vertices 
and all possible edges, the generation of all N$^{k-1}$MLT
is straightforward. 

In light of the compact and symmetric formulae of the MLT, NMLT and N$^2$MLT
configurations, we understood a pattern to generate causal representation of 
multi-loop topologies without making use of the LTD formalism. 
Following this alternative procedure, we found the causal representation
of all topologies up-to N$^7$MLT,
leaving, thus, a conjecture and procedure to generate any topology independently 
of the number of vertices.

We would be interested to see whether the approach studied in Ref.~\cite{Sborlini:2021owe}
could efficiently generate similar results to the ones presented in this paper. 
Given the fact that a purely geometrical approach was applied there, 
a connection with the current algebraic description 
will be desirable. 

\acknowledgments{
We are indebted to Germ\'an Rodrigo with whom the novel application of the 
loop-tree duality started and for numerous insightful discussions.
We also wish to thank German Sborlini for making available his results prior to publication
and for comments on the manuscript. 
This work is supported by the COST Action CA16201 PARTICLEFACE.
}

\bibliographystyle{JHEP}
\bibliography{refs}
\end{document}